\documentclass[aps,prb,twocolumn,superscriptaddress]{revtex4-1}
\pdfoutput=1
\usepackage{graphicx}
\usepackage{epstopdf}
\usepackage{array}
\usepackage{amssymb,amsmath}
\usepackage[centerlast,small]{caption2}

\newcolumntype{C}{>{$}c<{$}}
\newcolumntype{L}{>{$}l<{$}}
\newcolumntype{R}{>{$}r<{$}}

\renewcommand{\vec}[1]{\mbox{\mathversion{bold}$#1$}}

\begin{document}
\title{Band Gaps in Jagged and Straight Graphene Nanoribbons Tunable by an External Electric Field}
\author{V.A. Saroka}
\email{v.saroka@exeter.ac.uk}
\affiliation{School of Physics, University of Exeter, Stocker Road, Exeter EX4 4QL, United Kingdom}
\affiliation{Institute for Nuclear Problems, Belarusian State University, Bobruiskaya 11, 220030 Minsk, Belarus}

\author{K.G. Batrakov}
\affiliation{Institute for Nuclear Problems, Belarusian State University, Bobruiskaya 11, 220030 Minsk, Belarus}

\author{V.A. Demin}
\affiliation{Emanuel Institute of Biochemical Physics of Russian Academy of Sciences,  Kosygina 4, 119334 Moscow, Russia}

\author{L.A. Chernozatonskii}
\affiliation{Emanuel Institute of Biochemical Physics of Russian Academy of Sciences,  Kosygina 4, 119334 Moscow, Russia}

\begin{abstract}Band gap control by an external field is useful in various optical, infrared and THz applications. However, widely tunable band gaps are still not practical due to variety of reasons. Using the orthogonal tight-binding method for $\pi$-electrons, we have investigated the effect of the external electric field on a subclass of monolayer chevron-type graphene nanoribbons that can be referred to as jagged graphene nanoribbons. A classification of such ribbons was proposed and band gaps for applied fields up to the SiO$_2$ breakdown strength ($1$ V/nm) were calculated. According to the tight-binding model, band gap opening (or closing) takes place for some type of jagged graphene nanoribbons in the external electric field that lays in the plane of the structure and perpendicular to its longitudinal axis. Tunability of the band gap up to $0.6$ eV is attainable for narrow ribbons. In the case of jagged ribbons with armchair edges larger jags forming a chevron pattern of the ribbon enhance the controllability of the band gap. For jagged ribbons with zigzag and armchair edges regions of linear and quadratic dependence of the band gap on the external electric field can be found that are useful in devices with controllable modulation of the band gap.
\end{abstract}

\maketitle
\section{Introduction}
Since it was obtained in a freestanding form graphene\cite{Novoselov2004} has been attracting the attention of the scientific community both as an interesting object for fundamental study (due to its massless Dirac fermions\cite{Novoselov2005}) and as a base for future technological advances, due to its chemical stability, mechanical strength\cite{Lee2012}, high electrical and thermal\cite{Balandin2008,Chen2010} conductivities. Numerous graphene applications\cite{Chernozatonskii2014} include field effect transistors\cite{Schwierz2010} and their interconnects\cite{Murali2009a,Interconnects2013}, sensors\cite{Schedin2007,Bi2013}, hydrogen storage\cite{Tozzini2013}, terahertz emitters\cite{Mikhailov2012,Batrakov2012}, transparent electrodes etc. 
Many of these applications would benefit from a full control over the band gap. For instance, great efforts have been undertaken to develop novel tunable sources and detectors of THz radiation\cite{Hartmann2014}. Speaking of electronic applications, one must confess that there is a problem of contact resistance\cite{Smith2013} that can be easily overcome in all carbon electronic devices\cite{Liang2010}, but this again requires a complete control over the band gaps of the nanostructures. Additionally, all carbon electronics can be easily recycled and used in a closed-loop production cycle. Therefore it is desirable to control the band gap both by structural modification and external fields.

A number of techniques have been proposed for band gap engineering in graphene: patterning of graphene \cite{Dvorak2013,Chernozatonskii2007}, straining of graphene \cite{Choi2010,Li2010,Ribeiro2009,Pereira2009,Chernozatonskii2010}, lateral confinement of charge carries in one dimension in graphene nanoribbon (GNR), vertical inversion symmetry breaking in bilayer graphene\cite{McCann2006a,Castro2007} or trilayer graphene\cite{Lui}. All of them have some advantages and disadvantage, for instance, patterning allows higher current compared to GNR, while symmetry breaking by an external electric field provides tunability of the band gap, giving rise to tunable devices which is not the case for patterning or carrier confinement.

The universal approach to band gap control was also attempted in a combination of strategies. Recently, after extensive theoretical study\cite{Sahu2010a,Sahu2008}, simultaneous lateral carrier confinement and vertical inversion symmetry breaking in bilayer GNRs to tune their band gaps and improve the on/off ratio in logic devices has been experimentally verified and reported\cite{Yu2013}.

However, as was noticed earlier in the literature\cite{Chang2006,Son2006,Huang2008}, owing to the presence of edges for GNR electric field can have not only an out-of-plane direction that is normal to the ribbon plane, but also an in-plane one that is transverse to the ribbon longitudinal axis. A crucial difference between these two geometries of applied field is that the latter may affect properties of one-layer systems. Even though width dependence was reported and band gap increase for zigzag GNR width decrease was shown\cite{Chang2006,Huang2008}, quite wide ribbons were chosen for the sample study and the limit of this increase for narrow (sub-$5$nm) GNR was not demonstrated. Probably, it could be explained by the natural desire to be closer to experimentalists who were mostly limited at that time by electron beam lithography and etching techniques of GNR production. Taking into account the latest advances in GNR synthesis, when GNRs with atomically precise edges were synthesised and a bottom-up approach to synthesis of chevron-type GNRs was demonstated\cite{Cai2010}, one must feel free to consider narrow GNRs. It is worth noticing that considerable steps forward have been made in the technique scalability and processability. At the moment it is possible to synthesize over $1$ gram of ribbons in a single synthesis \cite{Vo2014}. Quite long ($>200$ nm) ribbons can be synthesized and modified in solution\cite{Narita2014} that allows their deposition on any conductive or isolative substrate. Recently heterojunctions in chevron-type nanoribbons were also produced and investigated\cite{Cai2014}. All of these experimental efforts have resulted in the introduction of a new class of carbon nanostructures of a chevron type and their derivatives which are referred to by some authors as graphene nanowiggles. These structures have been attracting much attention in last years\cite{CostaGirao2011,Girao2012,Huang2011,Liang2013a,Liang2012,Liang2013,Chen2010}, although their history can be traced back to the pioneering theoretical works \cite{Wu2008,Sevincli2008b}.

Here we investigate an effect of the in-plane homogeneous electric field for a subclass of chevron-type GNRs and compare it with the effect for ordinary GNRs having straight edges, e.g. GNRs of zigzag and armchair types.
In Section \ref{Str} we introduce the structures under consideration and discuss their similarities and differences from those in other recent works\cite{CostaGirao2011,Girao2012,Huang2011,Liang2013a,Liang2012,Liang2013}. We also discuss the connection of their description with that of straight GNRs. Section \ref{Meth} is devoted to the method used. In Section \ref{Res}  we report on our obtained results. Finally, in Section \ref{Conc} we summarize our findings and present our conclusions.

\section{\label{Str} Structures}
In what follows we will refer to structures being studied as jagged graphene nanoribbons (JGNRs). The main novelty of JGNR is so-called jags introduced in Ref.\cite{Ihnatsenka2009}. However, in our case jags are placed asymmetrically on both side of the ribbon so that the mirror reflection symmetry with respect to the ribbon longitudinal axis is broken. Breaking of the reflection symmetry admits the existence of very narrow ribbons of such class without a host ribbon. They can be referred to as assembled graphene nanoribbons, if one wants to highlight the fact that they can be represented as a sequence of ordinary GNRs with straight edges connected at a specific angle that can be referred to as an apex one. In this regard JGNRs are like GNWs in papers\cite{CostaGirao2011,Girao2012,Huang2011,Liang2013a,Liang2012,Liang2013}. However, GNWs can have a truncated jag form, while jag arms are always of equal lengths so that they are reflection symmetric with respect to the axis that is orthogonal to the longitudinal one. Conversely, JGNR edge form is always zigzag, whereas jag arms can be of different lengths. Zigzag just mentioned must not be confused with zigzag type of edge.

\begin{figure*}
\begin{center}
\includegraphics[scale=0.8]{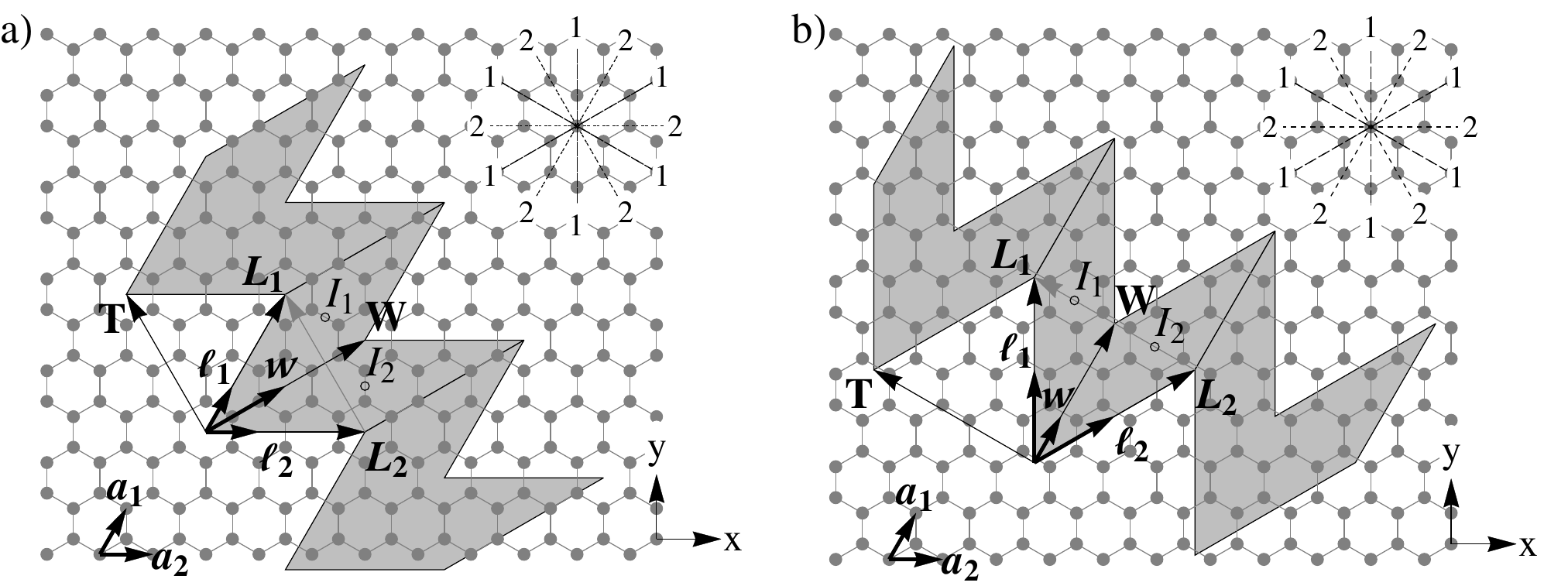}
\end{center}
\caption{\label{Characteristic_vectors} JGNRs characteristic vectors for zigzag a)  and armchair b) ribbon orientations on a graphene sheet represented by ribbons Z60$\langle 3,3;2 \rangle$ and A60$\langle 2,2;3 \rangle$, respectively. Lines 2-2 and 1-1 (insets) show all possible zigzag (Z) and armchair (A) directions, respectively. $\vec{L}_1$, $\vec{L}_2$ and $\vec{W}$ are chief vectors, $\vec{\ell}_1$, $\vec{\ell}_2$ and $\vec{w}$ are elementary vectors,  $\vec{a}_1$ and $\vec{a}_2$ are primitive translations of the graphene lattice, $\vec{T}$ is the JGNR translation vector and $I_1$,$I_2$ are the centres of JGNR inversion symmetry.}
\end{figure*}

To describe a jag one needs two vectors $\vec{L}_1$ and $\vec{L}_2$. Their magnitudes correspond to the jag arm lengths, while the angle between them is the apex angle. In one respect, JGNR is even more simple than GNR with symmetric positioning of jags\cite{Ihnatsenka2009} because one edge of JGNR can be obtained from another by mere translation on a certain vector, which is reasonable to determine as width vector $\vec{W}$ (See Fig.\ref{Characteristic_vectors}).
These three vectors describing the macrostructure of the superlattice are the chief vectors of JGNR, but they are still not connected with  the microstructure of JGNR, e.g. there is no reference to the graphene hexagon lattice. In the most general case they could be any lattice vectors having the form
\begin{equation}
\vec{C} = n \vec{a}_1 + m \vec{a}_2;
\label{arbitrary_lattice_vector}
\end{equation}
where $\vec{a}_1$ and $\vec{a}_2$ are the primitive translations of the graphene lattice. Introducing the greatest common divisor ($d$) of $n$ and $m$, one can rewrite Eq.(\ref{arbitrary_lattice_vector}) in the form
\begin{equation}
\vec{C} = d \left(v_1 \vec{a}_1 + v_2 \vec{a}_2\right) = d \vec{v};
\end{equation}
where $\vec{v}$ can be referred to as an elementary vector for the specified direction as $v_1$ and $v_2$ do not have common divisor except for unity. For two particular types of directions in the graphene lattice that are referred to as zigzag (Z) and armchair (A), the elementary vectors are the smallest possible having magnitudes of $\sqrt{3} a_0$ and $3 a_0$, respectively, where $a_0$ is the distance between the nearest carbon atoms. Therefore, in this work we restrict the chief vectors $\vec{L}_1$, $\vec{L}_2$ and $\vec{W}$ only to these directions depicted by 2-2 and 1-1 lines in the Fig.\ref{Characteristic_vectors}.

\begin{figure*}
\begin{center}
\includegraphics[scale=0.8]{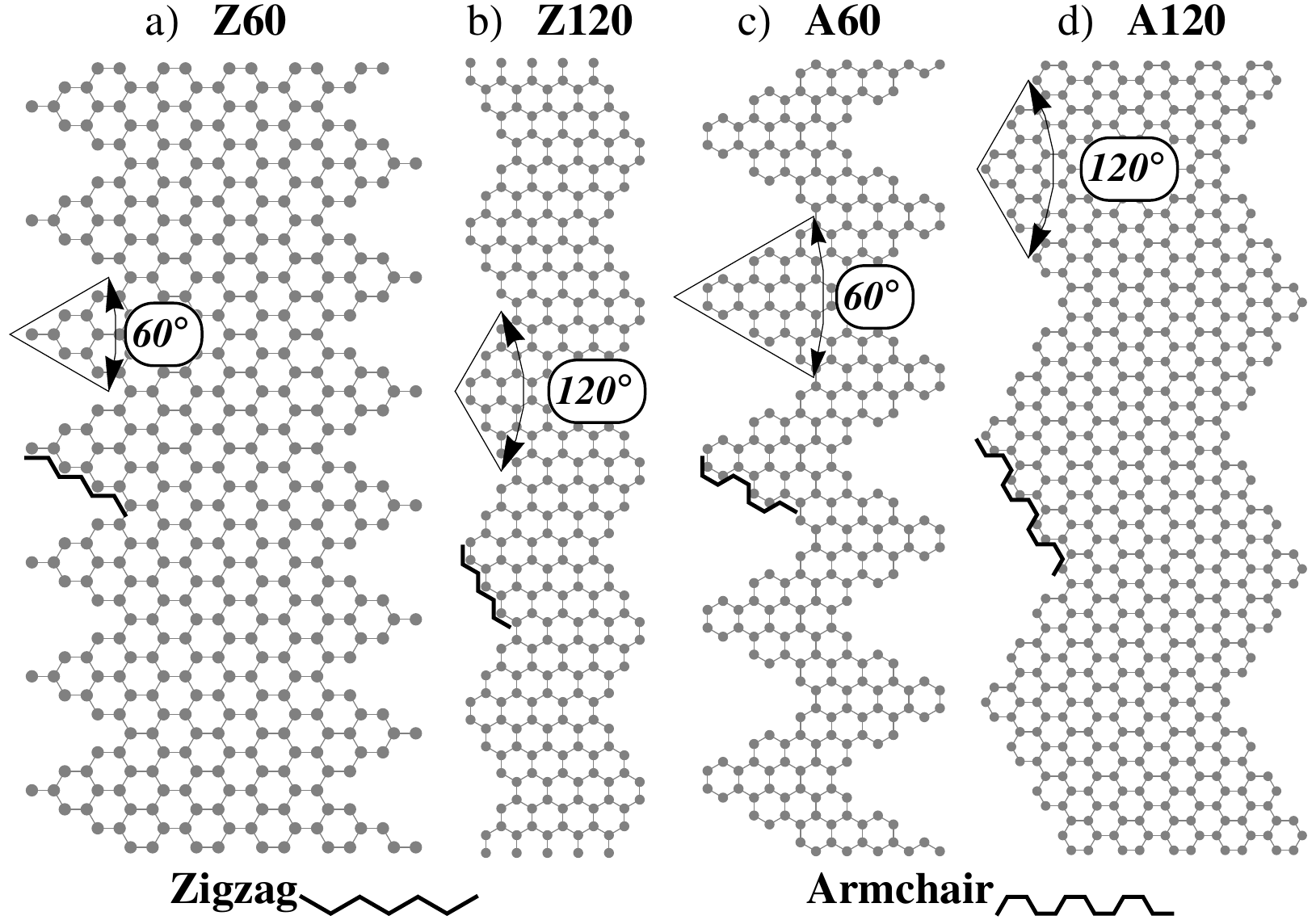}
\end{center}
\caption{\label{JGNR types} JGNRs types represented by ribbons Z60$\langle 3,3;5 \rangle$, Z120$\langle  3,3;5 \rangle$, A60$\langle  3,3;5 \rangle$ and A120$\langle  3,3;5 \rangle$, respectively.}
\end{figure*}

As a result, one can label any JGNR by a set of three integers corresponding to the chief vectors $\vec{L}_1$, $\vec{L}_2$ and $\vec{W}$. These integer indexes uniquely define JGNR if the set of elementary vectors is known (see Table \ref{Elementary_vectors}).
\begin{table}
\caption{The coordinates of JGNR elementary vectors $\vec{\ell}_1$, $\vec{\ell}_2$ and $\vec{w}$ in the basis of primitive translations of graphene lattice $\vec{a}_1$ and $\vec{a}_2$. }\label{Elementary_vectors}
\begin{ruledtabular}
\begin{tabular}{lcccc}
 & Z60 & Z120 & A60 & A120 \\ \hline
$\vec{\ell}_1$ & $(1,0)$ & $(1,-1)$ & $(2,-1)$ & $(2,-1)$ \\
$\vec{\ell}_2$ & $(0,1)$ & $(0,1)$ & $(1,1)$ & $(-1,2)$ \\
$\vec{w}$ & $(1,1)$ & $(1,0)$ & $(1,0)$ & $(1,1)$ \\ 
\end{tabular}
\end{ruledtabular}
\end{table}
For one set of elementary vectors there is a manifold of index sets, therefore the set of elementary vectors can be considered as characteristic of a JGNR type. But the set of vectors are not so convenient for structure type labelling as an index set for ribbon one. Therefore, taking into account that structures corresponding to various vector sets differ from one another only by their edge types and apex angles, we propose the following notation "t"$ \varphi$, where "t" stands for "type" and $\varphi$ -- for the apex angle. If one demands the width vector to be a bisector of the apex angle, then both jag arms have the same type of edges, e.g. zigzag or armchair, and can be denoted as Z or A. Thus, all variety of JGNR is reduced to four types Z60, Z120, A60, A120 and within each type any ribbon is specified by index set $\langle \ell_1,\ell_2;w \rangle$ (see Fig.\ref{JGNR types}).

\begin{figure*}
\begin{center}
\includegraphics[scale=0.8]{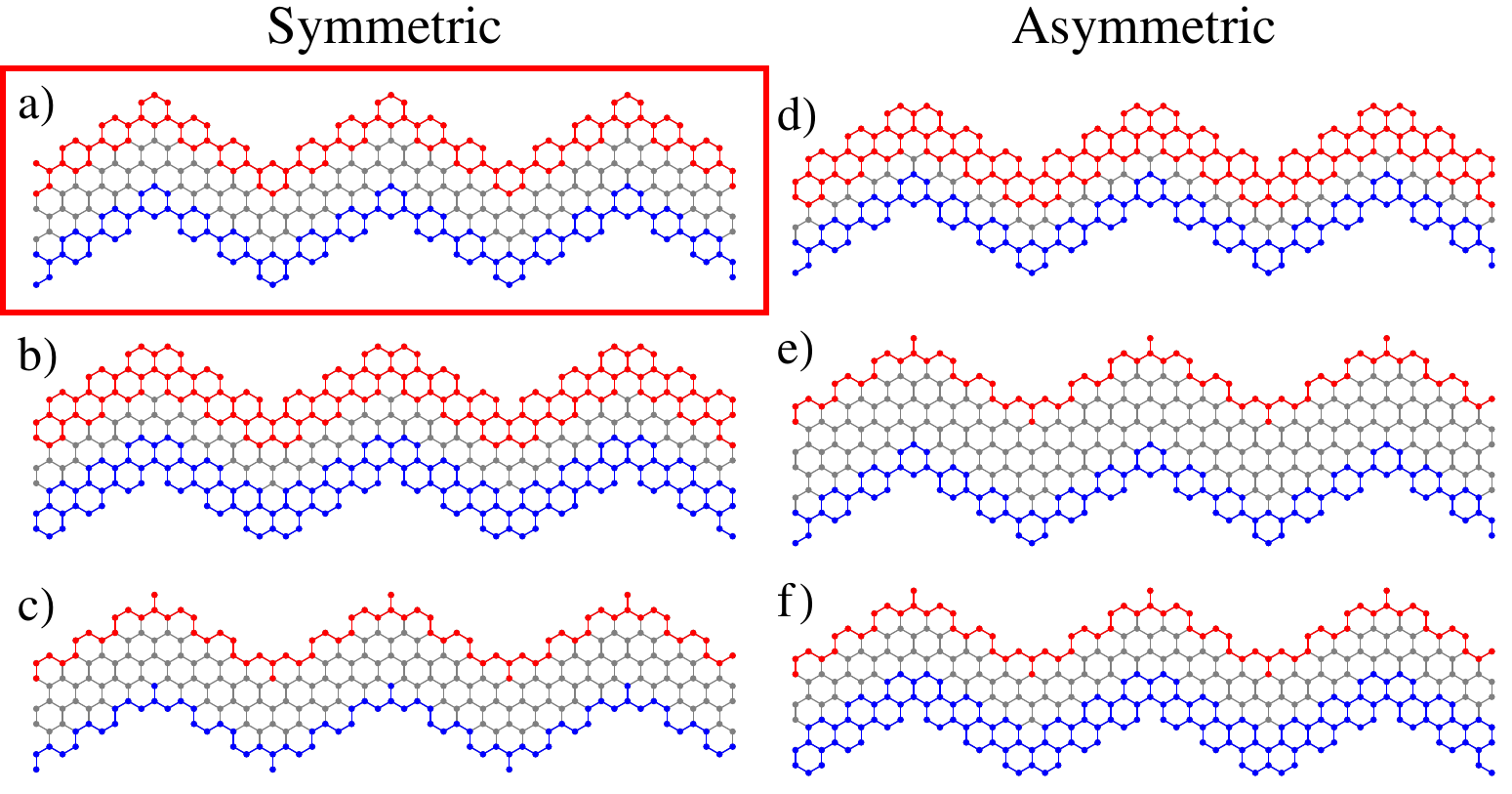}
\end{center}
\caption{\label{A120} JGNRs A120 subtypes. Ribbons shown on a), b), c) are symmetric, because their opposite edges highlighted by red (top) and blue (bottom) can be superimposed by simple translation, e.g. they have inversion centres like $I_1$,$I_2$ in Fig.\ref{Characteristic_vectors}. Ribbons depicted on d), e), f) are asymmetric due to different sorts of their opposite edges. Only a) and f) ribbons can be smoothly rolled into a nanotube. The framed ribbon is of the same A120 type as the one shown in Fig.\ref{JGNR types} d).}
\end{figure*}

By means of the introduced vectors and indexes such quantities as JGNR superlattice translation vector $\vec{T}$ and the number of atoms in the unit cell $N$ can be expressed:
\begin{eqnarray}
\vec{T} & = & \vec{L}_1 - \vec{L}_2; \label{Translation} \\
N & = & \lambda w (\ell_1 + \ell_2);
\end{eqnarray}
where $\lambda = 2$ in case of JGNR Z60, Z120, A60, and $\lambda = 6$ in case of JGNR A120. The difference in values of $\lambda$ arises due to our intention to preserve inversion symmetry (There are two centres of such symmetry for each JGNR, see Fig.\ref{Characteristic_vectors}) of the structures. A smaller step in the A-direction to increment width or smaller elementary vector for width can be determined, but then inversion symmetry on microstructure level for JGNR will be lost, as shown in Fig.\ref{A120}. As one can see, there are three possible subtypes of JGNR A120 symmetric with respect to inversion operation, but only one of them, see Fig.\ref{A120} a), can be rolled into a nanotube like JGNRs of other types and only such ribbons will be considered further.

\begin{figure*}
\begin{center}
\includegraphics[scale=0.8]{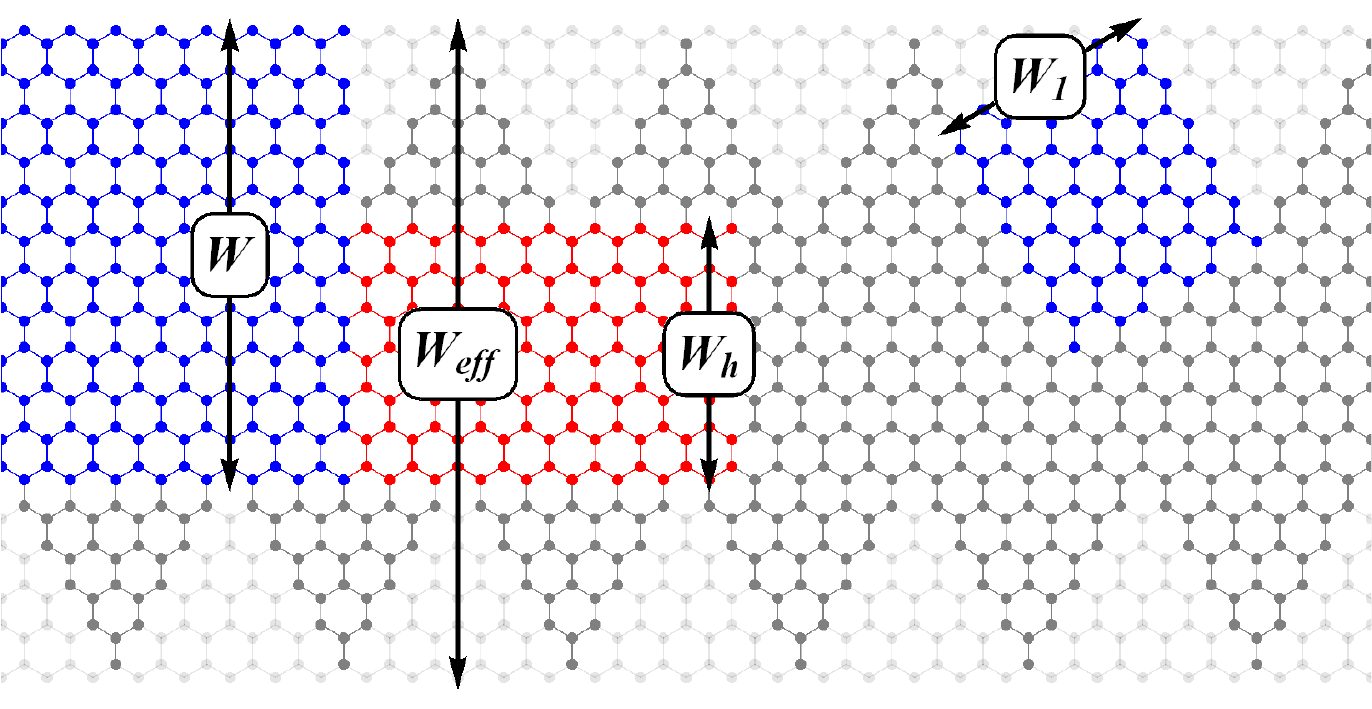}
\end{center}
\caption{\label{Widths} JGNR Z60 $\langle 5,5;6 \rangle$ mapping onto straight GNRs, where $W$ -- JGNR and the straight ribbon width, $W_{eff}$ -- the effective ribbon width, $W_h$ -- the host ribbon width, $W_1$ -- the arm ribbon width.}
\end{figure*}

So far, only two points of view on the structure of JGNR were mentioned. On one hand, it can be represented as a sequence of ordinary straight GNR pieces concatenated at specific angles, but on the other hand it can be considered as an ordinary straight GNR with some triangle areas cut to produce jags, but there are some more of them, see Fig.\ref{Widths}.
Therefore the width  introduced above is not so a complete characteristic. However, on its basis one can determine others. For instance, an effective width, e.g. the width of the GNR that one cut some parts from to produce jagged edges, can be expressed as follows:
\begin{equation}
\vec{W}_{eff}  = \frac{ \vec{T} \times \vec{P}_{1(2)} \times \vec{T}}{\vec{T}^2}\, ,
\label{Effective_width}
\end{equation}
where $\vec{P}_{1(2)}  =  \vec{L}_{1(2)} + \vec{W}$. This definition is independent on choice of vector $\vec{L}_1$ or $\vec{L}_2$ and is useful for result comparison to clarify the role of jags in electronic properties of JGNR. Other characteristics closely related to the width are widths of the JGNR arm ribbons which have the form of
\begin{equation}
\vec{W}_{1(2)}  =  \frac{ \vec{L}_{1(2)} \times \vec{W} \times \vec{L}_{1(2)}}{\vec{L}_{1(2)}^2}\, .
\label{Arm_width}
\end{equation}
For the present paper, as the width vector was chosen to be the bisector of the apex angle, these two quantities defined by Eq.(\ref{Arm_width}) are equivalent. Finally, one can think about some JGNRs as a host ribbon with attached triangle fragments on both sides (see Fig.\ref{Widths}), therefore for quite wide JGNR one can introduce a host ribbon width:
\begin{equation}
\vec{W}_{h}  = 2 \vec{W} - \vec{W}_{eff}\,.
\label{Host_width}
\end{equation}
Equating the right hand side of Eq.(\ref{Host_width}) to zero, one can easily obtain the criterion for the host ribbon presence in the structure for each type of JGNRs: Z60 $\ell \geq 2w$, Z120 $\ell \geq 2 w$, A60 $\ell > 2 w/3$, A120 $\ell \geq 2w$. Such simple mapping of JGNR onto ordinary straight GNRs is only possible in symmetric case when $\ell_1 = \ell_2$. As for the asymmetric case, it is more complicated, because the effective, the host and the straight ribbons become specific JGNRs with low values of indexes. Additionally, this mapping would be of no practical use for ribbons from the same class that can be compared straightforwardly.

\section{\label{Meth} Method}
The electronic structure of the presented JGNRs was investigated within the orthogonal tight-binding model for $\pi$-electrons. Electronic bands were obtained as eigenvalues for the matrix Hamiltonian $H$ with elements of the following form:
\begin{equation}
H_{nn^\prime} = \sum_{i,j,q}  t_{ni} \exp \left(i \vec{k} \vec{r}_{nij} \right) \delta_{ \scriptsize \vec{r}_{n^{\vphantom{\prime}}ij}, \vec{r}_{n^{\vphantom{\prime}}1} - \vec{r}_{n^\prime q}} \, \label{matrixHamiltonian}
\end{equation}
where $\vec{r}_{nij}$ is a vector pointing the $j$-th position of the $i$-th order  nearest neighbour from the $n$-th atom position in the unit cell of the structure, $\vec{r}_{nq}$ is a radius-vector pointing the position of the $n$-th atom in the $q$-th unit cell.
Within the tight-binding model, the so-called hopping integral $t_{ni}$ for $n$-th atom in the unit cell of the structure is expressed as
\begin{equation}
t_{ni} = \langle \phi_n | \hat{H} | \phi_{n+i} \rangle\, ,
\end{equation}
where $\hat{H}$ is a system Hamiltonian, $\phi_n$ is an atomic orbital of the $n$-th atom and $i$ can be referred to as the neighbour order. 
In the presence of an external field the Hamiltonian takes form
\begin{equation}
\hat{H} = \hat{H}_0 + \hat{U}\, ,
\end{equation}
where $\hat{H}_0$ is the Hamiltonian of the system without external field, $\hat{U}=-e\vec{E}\vec{r}$ is the potential energy operator for a homogeneous electric field. Taking this into account one has $t_{ni}=t_{ni,0} + \delta t_{ni}$, with $\delta t_{ni} = \langle \phi_n | \hat{U} | \phi_{n+i} \rangle $. Note that the external electric field is much less than the atomic one, i.e.,
\begin{equation}
eE \ll  \frac{t_{n1,0}}{a_0} \, ,
\label{field_crit}
\end{equation}
where $t_{n1,0} \approx 3$ eV, $a_0=1.42\mathring{\mbox{A}}$.
This allows one to neglect any change in the atomic orbitals $\phi_{n},\phi_{n+i}$ due to the field. It is also worth mentioning that we do not expect large field enhancement at the sharp ends of jags as their sizes in the structures considered are about $5$nm, which is quite small, and they are arranged periodically in an infinite line so that their influence on each other averages and reduces the resulting field.

As we are interested in the pure effect of the jagged edges and its influence on electronic properties of JGNRs, we eliminate possible differences of hopping integrals at various sites within unit cell so that $t_{ni} = t_i;\, t_{ni,0} = t_{i,0} ; \, \delta t_{ni} = \delta t_i$. In fact, the model of ideal geometry was implemented, which is close to the real geometry of the structure when dangling sp$^2$ bonds at JGNR edges are passivated with H atoms to form a strong C-H $\sigma$-bonds. Nearest-neighbor C-H matrix elements associated with these H-atoms do not appear explicitly in $H_0$ because these $\sigma$ orbitals are symmetric with respect to the nodal plane of the $\pi$-orbitals and hence decouple from the $\pi$-conduction bands described by $H_0$. Also, because C forms a strong covalent bond with H, the $\sigma$ and $\sigma^{\ast}$ bands associated with these bonds lie far from $E_F$ and hence need not be considered further.

It is worth noting that on one hand hopping integral values at the edge can differ from those in the GNR interior so that it could cause band gap opening in armchair GNR\cite{PhysRevLett.97.216803}. But on the other hand, the same band gap opening can be explained by accounting for higher order hopping integrals\cite{White2007}. A more accurate model taking into account both effects was reported latter\cite{Gunlycke2008}. Interestingly, a more precise model accounting for edge distortions showed that this effect leads to a small correction only for armchair ribbons of $3n$ series. However, it is very unlikely that this small correction to the band gap does affect its tunability. Therefore, admitting an error for the band gap of no more than $10$\%,  the nearest neighbours up to the third order were taken into account and model parameters were chosen as proposed in the paper\cite{White2007}: $t_{0,0} = 0$ eV, $t_{1,0} = -3.2$ eV, $t_{2,0} = 0$ eV $t_{3,0} = -0.3$ eV. 

In the present paper we neglect screening effect and assume that applied field influences directly each lattice site. It seems reasonable for strong electric fields. However, for weak fields one should keep in mind that the values presented should be interpreted as a value of an effective field.
The applied electric field contributes mainly to the site energy $ t_{0}$. In fact, there are additional terms for higher order hopping integrals $\delta t_{1 \ldots \infty}$. To assess them one can  approximate carbon $\pi$-orbitals by $p_z$ wave functions of the hydrogen like atom\cite{Partoens2006}:
\begin{widetext}
\begin{equation}
\phi_n( |\vec{r} - \vec{r}_n|, \theta ) =\frac{1}{4 \sqrt{2 \pi }} \left(\frac{Z}{a_B} \right)^{5/2} \exp \left(-\frac{ Z |\vec{r} - \vec{r}_n| }{2 a_B } \right) |\vec{r} - \vec{r}_n| \cos \theta,
\end{equation}
\end{widetext}
where $Z=6$ (for carbon), $a_B$ is the Borh radius and $\theta$ is the polar angle measured from the $z$-direction (in spherical coordinate system), and calculate the ratios $\delta t_{i}/ t_{i,0}$. Obviously, the value obtained for $i = 0$ is of greater significance than any other. That is why we adopted $\delta t_{1 \ldots \infty} \approx 0$, $\delta t_{0} = - \vec{E} \vec{r}$ eV.

The band gap dependence on applied electric field was investigated for electric strength magnitudes up to the $0.1$ V/$\mathring{\mbox{A}}$ $ = 1$ V/nm. This value is $10$ times less than typical strength used in field emission calculations\cite{Tada2002,Araidai2004}, but corresponds to the breakdown strength of electric field in SiO$_2$\cite{DiMaria1993}. Even though it seems to be a quite high magnitude of field, it must be noticed that it still meets the requirement of Eq.(\ref{field_crit}). It is often chosen as a natural upper limit by others\cite{Castro2007,Sahu2008,Huang2008} for the electric field applied normally to the plane of the structure. Thus, our choice makes for an easier comparison of the two field arrangements. It is worth noting that break down values are very sensitive to the material used. For instance, in experimental work\cite{Yu2013} it was possible to increase the value more than twice by using a combination of SiO$_2$ with HfO$_2$, so that the resulting value was $\sim 0.033$ V/$\mathring{\mbox{A}}$, that is only $3$ times lower than mentioned above.

\section{\label{Res} Results}
\begin{figure*}
\begin{center}
\includegraphics[scale=0.7]{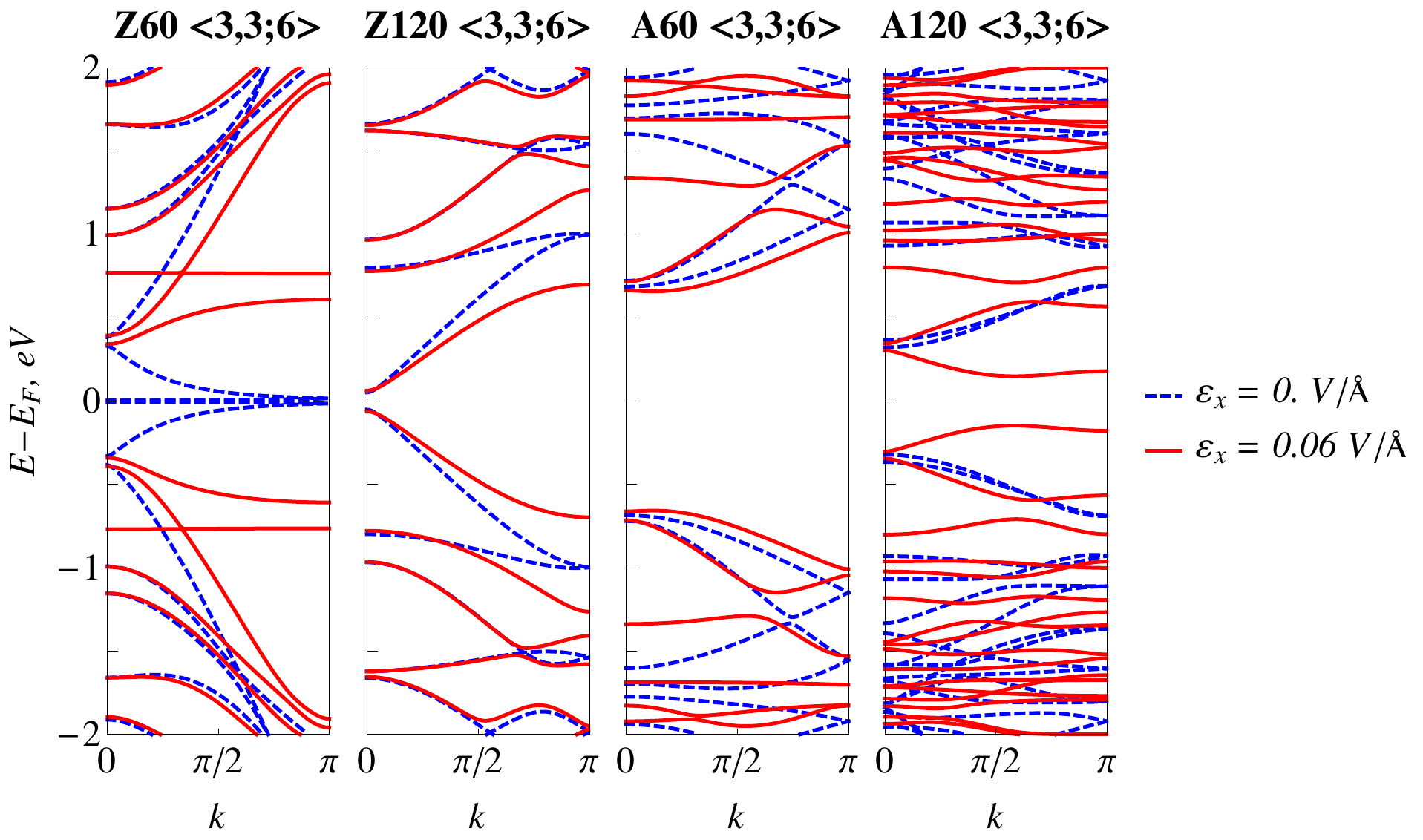}
\end{center}
\caption{\label{Field_Effect} The effect of a transverse electric field $\varepsilon_x$  on the band structure of four types of JGNRs, where $E_F$ -- the Fermi energy. $k$ is expressed in the units of $1/T$ everywhere.}
\end{figure*}

Let us now consider typical JGNR band structures and their changes in an external electric field. Throughout this paper all ribbons are assumed to be laying in the $xy$-plane and the $y$-axis is collinear with the ribbon translation vector so that transverse electric field strength has only one non-zero component $\varepsilon_x$. In Fig.\ref{Field_Effect}, one can see that the most profound changes take place for JGNR Z60 and A120. In the first case, one deals with a band gap opening while in the second case -- with a band gap closing. Surprisingly, there are no observable changes for band gaps of JGNR Z120 and A60. However, some alterations in Z120 and A60 band structures can be noticed. For the sake of the clarity of the discussion, let us take the following convention: all band are numbered by indexes $J$ so that the first conduction band corresponds to $J=1$ and first valence to $J=-1$ and so on. Then it can be pointed out that a profound splitting takes place at the edge of the Z120 Brillouin zone for bands $J=\pm 1$ and $J=\pm 2$, respectively. Similar splitting but of less magnitude and shifted towards the center of Brillouin zone is observed for A60 bands $J=\pm 2$ and $J=\pm 3$, correspondingly. With respect to the ribbon A60 one must confess that the least variations occur for bands $J=\pm 1$ compared to equivalent bands of other ribbons presented. It means that the effective mass of charge carriers changes only slightly due to the presence of a transverse electric field. In the case of Z120 ribbons the change of the effective mass for low doping concentrations should also be negligible. However, more significant changes take place for ribbons Z60 and A120. In both cases profound band bending can be observed for the magnitude of electric field $\varepsilon_x = 0.05 $~V/$\mathring{\mbox{A}}$. In general it can be seen that the influence of the field is stronger at the edge of Brillouin zone, as has already been shown in nanotube and nanohelix superlattice properties analysis, which is very relevant to a JGNR in the external electric field as it creates for JGNR's electron a periodic electrostatic potential similar to that discussed in papers\cite{Kibis2005a,Kibis2005,Batrakov2010}.

\begin{figure*}
\begin{center}
\includegraphics[scale=0.7]{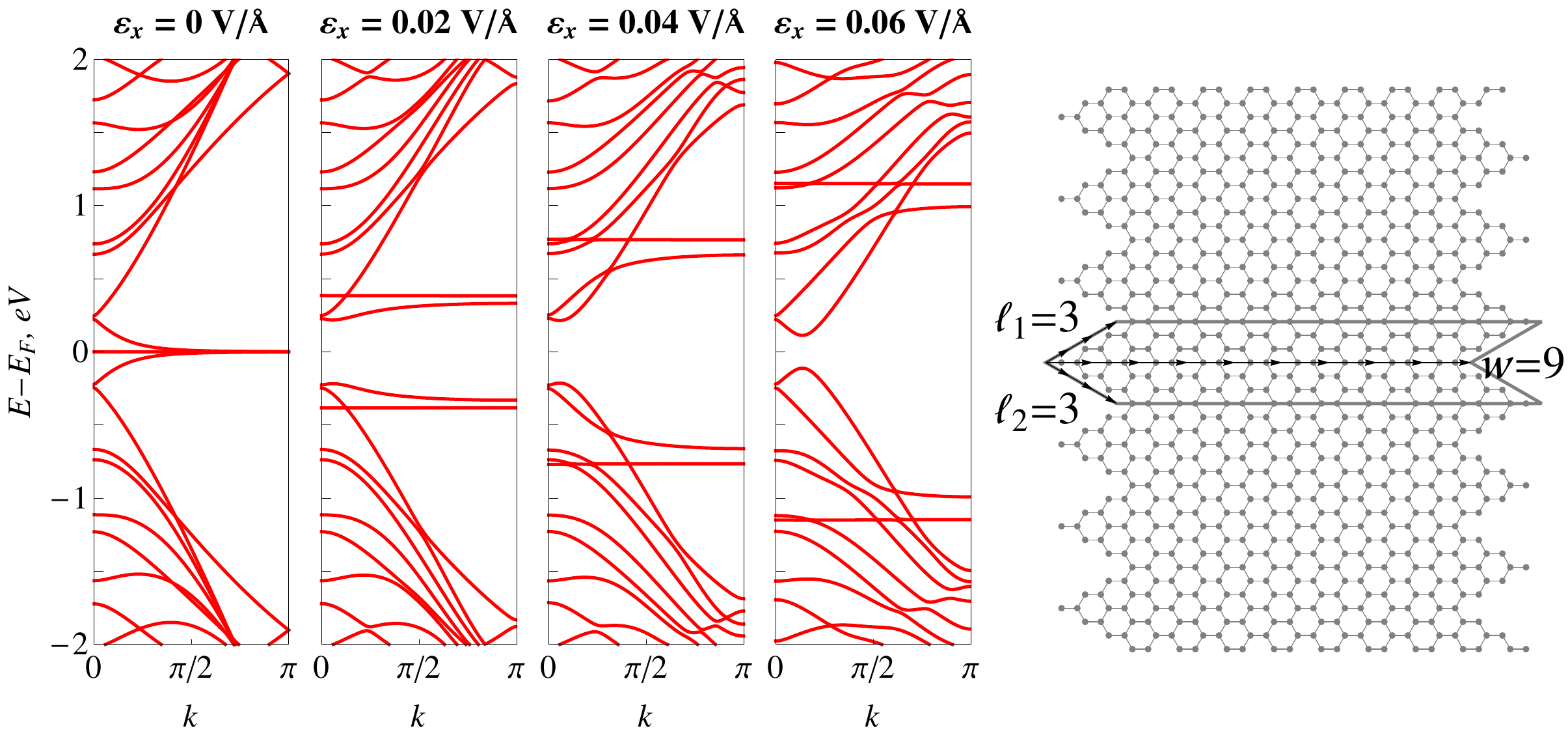}
\end{center}
\caption{\label{Field_Effect_Z60} The band structure evolution for increasing magnitude of transverse electric field for JGNRs Z60 $\langle 3,3;9\rangle$ (left) and atomic structure of the ribbon (right).}
\end{figure*}

\begin{figure*}
\begin{center}
\includegraphics[scale=0.7]{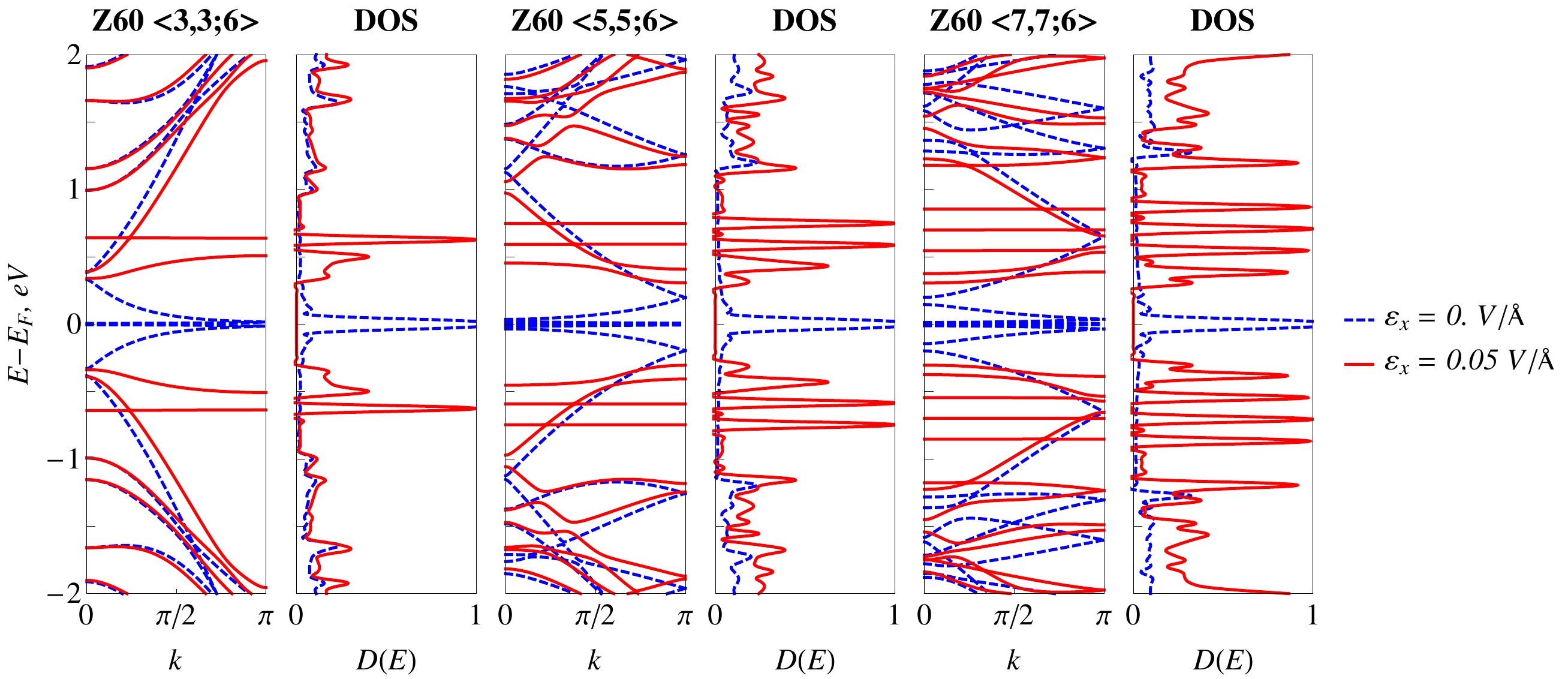}
\end{center}
\caption{\label{Splitting_Z60} The splitting of dispersionless bands in the transverse electric field for JGNRs Z60 $\langle 3,3;6\rangle, \langle 5,5;6 \rangle, \langle 7,7;6 \rangle $. DOS -- normalized density of states.}
\end{figure*}

Next we consider the band gap dependence on the magnitude of the transverse electric field in more detail. Let us proceed with the JGNR Z60. For Fig.\ref{Field_Effect_Z60} we took a bit wider ribbon Z60 $\langle 3,3;9 \rangle $ to show that for a high value of the electric field strength of about $0.06$~V/$\mathring{\mbox{A}}$  the band gap closing can be observed. The next feature that can be noticed in Fig.\ref{Field_Effect_Z60} is dispersionless bands, whose positions clearly correlate with the magnitude of the external field. These bands arise from zigzag edge states reported in the paper\cite{Fujita1996}. However, as JGNR is a superlattice with a longer translation period, its band structure if one does not take into account interactions leading to band anticrossing is a folded structure of straight GNR with zigzag edges (ZGNR). As one can see, due to such folding these bands are absolutely dispersionless through out the whole Brillouin zone compared to ZGNR where edge bands are dispersionless only though the $1/3$ of Brillouin zone. Changing the superlattice period of translation $\vec{T}$ by means of jag arms $\vec{L}_1$, $\vec{L}_2$ (see Eq.\ref{Translation}) one can control a number of foldings or put in other words a number of dispersionless bands. Without the external electric field almost all of them are degenerate, but with a switched on field their splitting is observed. These bands must result in sharp peaks in the density of states, whose positions in turn must be field dependent as well. However, as one can see from Fig.\ref{Field_Effect_Z60} for the high magnitude of electric strength of about $0.06$~V/$\mathring{\mbox{A}}$ they drown among multitude of other bands. But for narrower ribbons Z60 the density of states arising from all other bands except edge ones around the Fermi level is low in a quite wide region from $-1$ to $1$ eV, see Fig.\ref{Splitting_Z60}. It means that the peaks should act in this region as energy levels and we showed it for JGNR Z60 with $w=6$ in Fig.\ref{Splitting_Z60}. Taking into account all the just mentioned results, we predict that for JGNR Z60 with $w=4,..,8$  in an external transverse electric field, a series of electromagnetic emission and absorption lines will be observed and the number of lines will be consistent with $\vec{T}$. We tested that this result remains true for asymmetric JGNRs Z60 characterized by indexes $\ell_1 \neq \ell_2$ and for JGNRs Z60 of a slightly different structure, see the Appendix \ref{App}.

\begin{figure*}
\begin{center}
\includegraphics[scale=0.7]{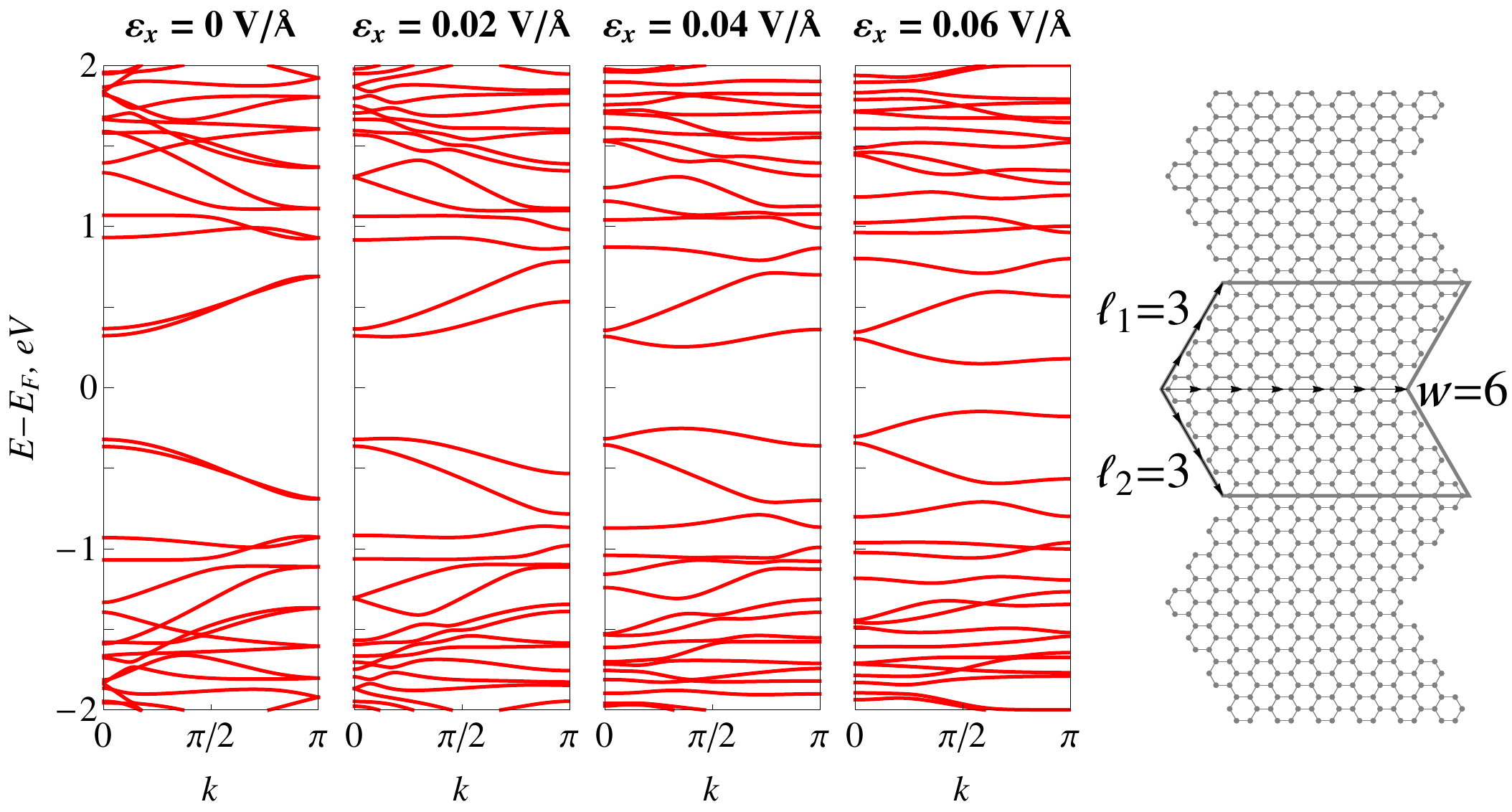}
\end{center}
\caption{\label{Field_Effect_A120} The band structure evolution for increasing magnitude of transverse electric field for JGNRs A120 $\langle 3,3;6\rangle$ (left) and atomic structure of the ribbon (right).}
\end{figure*}
Considering in Fig.\ref{Field_Effect_A120} the band gap closing for JGNR A120, one must notice that this closing takes place owing to the splitting of bands $J=\pm 1,\pm 2$ at the edge of the Brillouin zone. This splitting leads to the band gap shift in $k$-space from $k=0$ for $\varepsilon_x=0$~V/$\mathring{\mbox{A}}$ to the region $k > \pi/2$ for $\varepsilon_x=0.06$~V/$\mathring{\mbox{A}}$. Although there are some bands with low dispersion near energies $\pm 1$ eV their positions are not affected by the applied electric field. On the contrary, the almost flatness of the bands $J=\pm 1$ can be achieved for $\varepsilon_x=0.06$~V/$\mathring{\mbox{A}}$ in $k$-space from $k=\pi/2$ to $k=\pi$. It is very useful because it must increase the probability of interband transitions with frequency exactly corresponding to the band gap due to higher density of states on its both sides.
\begin{figure*}
\begin{center}
\includegraphics[scale=0.7]{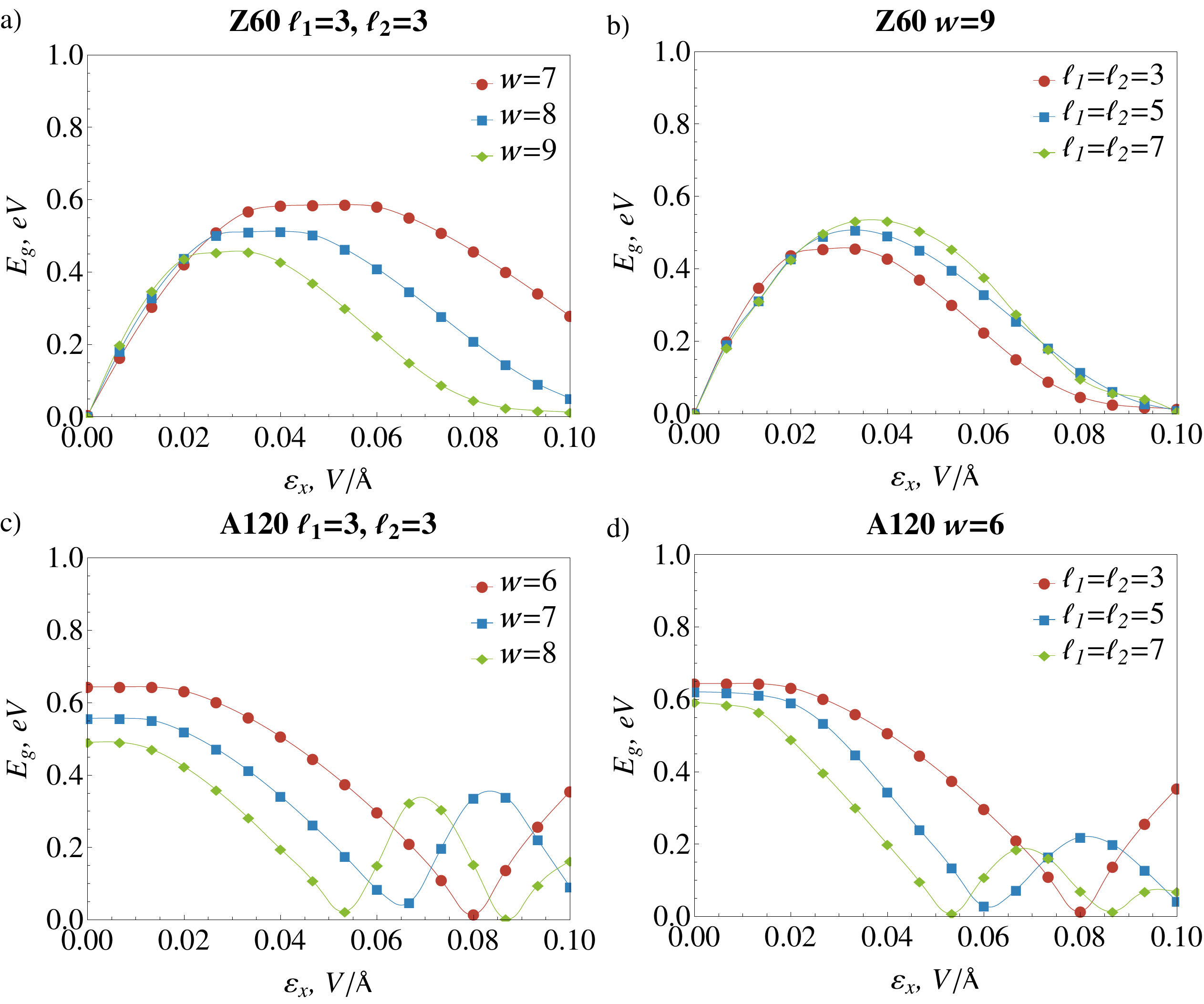}
\end{center}
\caption{\label{BandgapVsElectricfield} The band gap $E_g$ dependence on the magnitude of transverse electric field $\varepsilon_x$ for symmetric JGNR Z60 a), b) and A120 c), d) with various width and jag arm indexes $w$, $\ell_1 = \ell_2$, correspondingly.}
\end{figure*}
It is very interesting to see how the dependence of the band gap on transverse electric field is affected by symmetric JGNR parameters $w$ and $\ell_1 = \ell_2$. These results are presented for JGNRs Z60 and A120 in Fig.\ref{BandgapVsElectricfield}.
As it can be seen from Fig.\ref{BandgapVsElectricfield} a) the maximum value of the band gap opening $(\sim 0.6\mbox{ eV})$ is higher for the narrower ribbon with $w=7$ but it is attainable for stronger transverse electric field ($~0.04$~V/$\mathring{\mbox{A}}$) compared to Z60 $\langle 3,3;8\rangle$ and $\langle 3,3;9\rangle$, besides for the specified ribbon the band gap cannot be closed within the restricted range $<0.1$~V/$\mathring{\mbox{A}}$ as for ribbon Z60 $\langle 3,3;9\rangle$. For a low strength field $<0.02$~V/$\mathring{\mbox{A}}$ the band gap opening for narrower Z60 is a bit less than for wider ones, but it results in more significant divergence for greater difference of width indexes. The influence of jag arm index on the band gap opening of the ribbon Z60 shown in Fig.\ref{BandgapVsElectricfield} b) is opposite to what was just mentioned. For higher indexes the maximum value increases and drifts to greater values of $\varepsilon_x$ while band gap closing at $\varepsilon_x = 0.1$~V/$\mathring{\mbox{A}}$ is not affected by them. In low strength fields the difference in band gap opening is almost negligible.
According to Fig.\ref{BandgapVsElectricfield} c) the band gap closing does not take place immediately, and JGNRs A120 $\langle 3,3;6 \rangle$ are quite resistant to low electric fields with $\varepsilon_x < 0.02$~V/$\mathring{\mbox{A}}$. However this resistance decreases if the width index increases that leads to lower values of electric strength required to close the band gap completely: $\varepsilon_x = 0.055$~V/$\mathring{\mbox{A}}$ for A120 $\langle 3,3;8 \rangle$ and $\varepsilon_x = 0.08$~V/$\mathring{\mbox{A}}$ for A120 $\langle 3,3;6 \rangle$. Another fascinating feature of the ribbons  is the peak in the band gap dependence on electric field strength observed for A120 $\langle 3,3;7 \rangle$ at $\varepsilon_x = 0.85$~V/$\mathring{\mbox{A}}$. This peak position shifts to lower magnitudes of $\varepsilon_x$ as the width index $w$ increases while its height seems to remain unaffected. This is quite strange taking into account that there is an obvious decrease of initial values of the band gap, e.g. for $\varepsilon_x = 0$~V/$\mathring{\mbox{A}}$, for greater indexes $w$. Compared to this case the data presented in Fig.\ref{BandgapVsElectricfield} d) are very different. While the decrease of initial values of the band gap for incremented jag arm indexes $\ell_1$ and $\ell_2$ is less than in Fig.\ref{BandgapVsElectricfield} c) for increasing $w$ the resistance to the band gap closing is of the same measure, e.g. achieved at $\varepsilon_x = 0.053$~V/$\mathring{\mbox{A}}$ for A120 $\langle 7,7;6 \rangle$. This value is compared to the value for A120 $\langle 3,3;8 \rangle$, however in the case of A120 $\langle 7,7;6 \rangle$ one closes a wider band gap. In Fig.\ref{BandgapVsElectricfield} d) a peak similar to that one in Fig.\ref{BandgapVsElectricfield} c) in dependence of $E_g$ on $\varepsilon_x$ is manifested. Its position shifts to lower $\varepsilon_x$ for greater $\ell_{1}$, $\ell_{2}$ and it resembles behaviour of the peak in Fig.\ref{BandgapVsElectricfield} c) but its height decreases if jag arms indexes $\ell_{1}$,$\ell_{2}$ both increase. 

It is obvious that the dependences $E_g(\varepsilon_x)$ for Z60 and A120 JGNRs presented in Fig. \ref{BandgapVsElectricfield} cannot be described by simple functions, however they can be approximated by polynomials that are truncated Taylor series. Quite easily one can specify the regions where two or three terms of Taylor expansion are necessary for a reasonable approximation. These regions correspond to liner and quadratic dependences $E_g(\varepsilon_x)$ and a good knowledge about them is of great importance for possible application in linear and non-linear devices, therefore we calculated the parameters characterizing the rate of the band gap opening/closing and presented them in Table \ref{tab2fig9}. Moreover, Table \ref{tab2fig9} enables comparison between two types of ribbons, which is difficult to do by means of plotting. Although it might seem strange to compare different patterns we want to notice that ribbon A120 placed in sufficiently strong electric field so that the point of complete band gap closing, for instance $\varepsilon_x=0.08$~V/$\mathring{\mbox{A}}$ for A120 $\langle 3,3;6 \rangle$, is achieved can work in the same regime as Z60 ribbon --- metal--dielectric transition for decreasing field. As can be clearly seen from the Table \ref{tab2fig9} the absolute value of $\beta$ is larger for all Z60 than for A120 ribbons presented in the Table \ref{tab2fig9}, that means their band gaps can be controlled more efficiently.
\begin{table*}
\caption{Fitting of linear and quadratic regions of $E_g(\varepsilon_x)$ curves for JGNRs presented in Fig.\ref{BandgapVsElectricfield}. }\label{tab2fig9}
\begin{ruledtabular}
\begin{tabular}{cccccccccc}
 & \multicolumn{4}{c}{$\beta x + \alpha $} & \multicolumn{5}{c}{ $\gamma x^2+\beta x+ \alpha$} \\
 JGNR & $\alpha$, eV  & $\beta$, eV$\cdot$\AA/V  & $\varepsilon _{\min }$,  V/\AA & $\varepsilon
   _{\max }$,  V/\AA & $\alpha $, eV & $\beta $,  eV$\cdot$\AA/V & $\gamma $, eV$\cdot$\AA$^2$/V$^{2}$ & $ \varepsilon _{\min }$, V/\AA & $\varepsilon _{\max }$, V/\AA \\ \hline
 Z60 $\langle 3,3;7 \rangle$ & 0 & 23.3 & 0 & 0.010 & 0 & 29.8 & -361.4 & 0.010 & 0.040 \\
  Z60 $\langle 3,3;8 \rangle$  & 0 & 25.9 & 0 & 0.010 & 0 & 33.7 & -526.3 & 0.010 & 0.035 \\
  Z60 $\langle 3,3;9 \rangle$  & 0 & 29.6 & 0 & 0.007 & 0 & 35.7 & -680.7 & 0.007 & 0.030 \\
  Z60 $\langle 5,5;9 \rangle$  & 0 & 28.4 & 0 & 0.007 & 0 & 29.2 & -425.5 & 0.007 & 0.045 \\
  Z60 $\langle 7,7;9 \rangle$  & 0 & 27.0 & 0 & 0.007 & 0 & 27.7 & -365.0 & 0.007 & 0.055 \\
  A120 $\langle 3,3;6 \rangle$ & -- & -- & -- & -- & 0.7 & 0.6 & -108.4 & 0.010 & 0.080 \\
  A120 $\langle 3,3;7 \rangle$ & 0.8 & -12.5 & 0.035 & 0.060 & 0.6 & 1.4 & -177.8 & 0.010 & 0.035 \\
  A120 $\langle 3,3;8 \rangle$ & 0.7 & -12.5 & 0.025 & 0.050 & 0.5 & 2.1 & -264.4 & 0 & 0.025 \\
  A120 $\langle 5,5;6 \rangle$ & 1.0 & -15.7 & 0.035 & 0.060 & 0.6 & 8.0 & -349.5 & 0.010 & 0.035 \\
  A120 $\langle 7,7;6 \rangle$ & 0.8 & -14.8 & 0.020 & 0.050 & 0.5 & 9.5 & -616.2 & 0.007 & 0.020 \\
\end{tabular}
\end{ruledtabular}
\end{table*}

\begin{turnpage}
\begin{figure*}
\begin{center}
\includegraphics[scale=0.65]{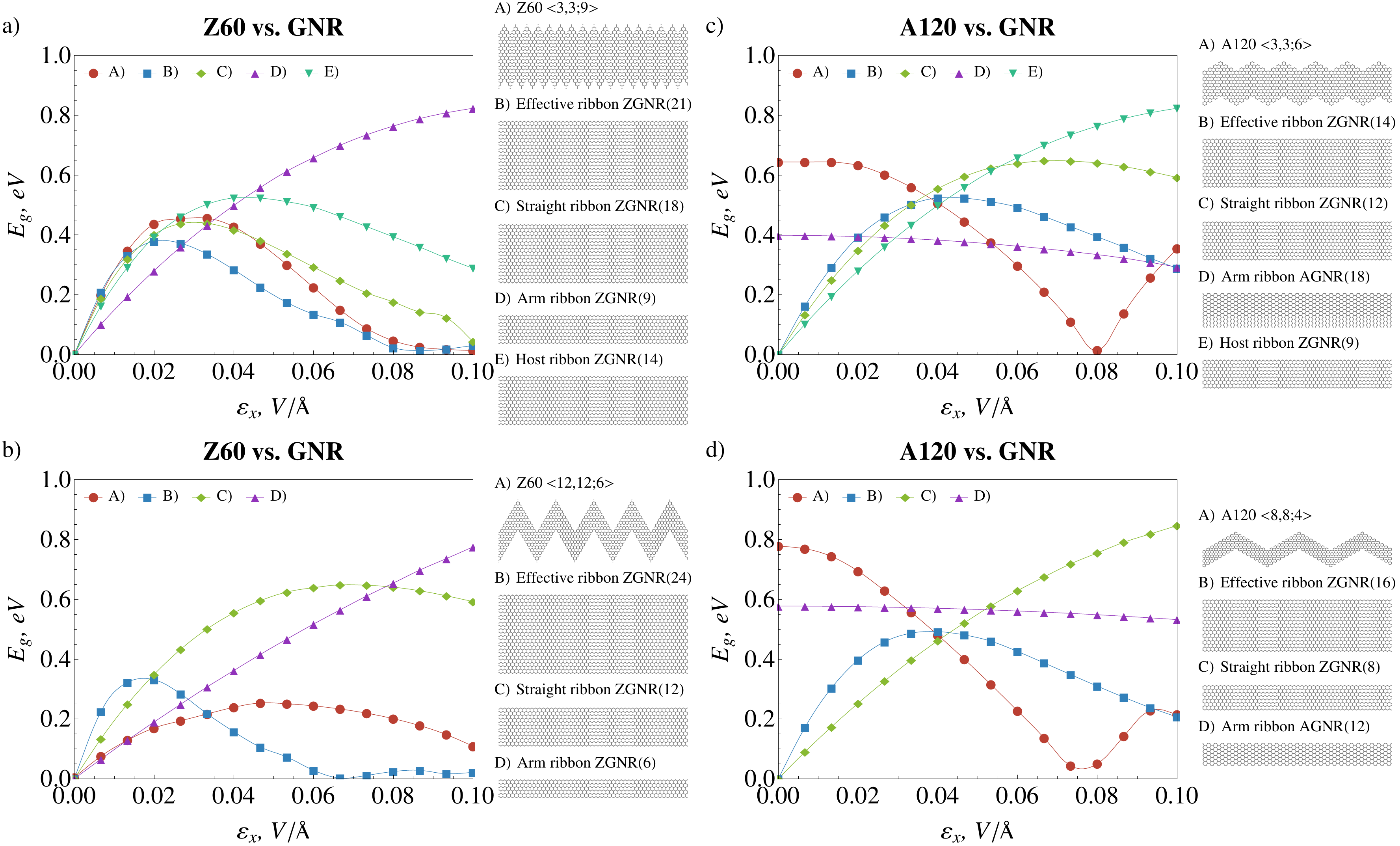}
\end{center}
\caption{\label{JGNRvsGNR} The comparison of JGNRs Z60 a), b) and A120 c), d) with various straight GNRs, where (n) denotes the number of atom pairs in the unit cell of the structure.}
\end{figure*}
\end{turnpage}
Having studied the dependence of $E_g$ on $\varepsilon_x$ for JGNRs Z60 and A120, we proceed with a comparison of found patterns with similar ones for simple GNR with zigzag and armchair edges, which will be referred to as ZGNR($n$) and AGNR($n$), respectively, where $n$ is a number of carbon atom pairs in the unit cell of the structure. The mapping mentioned at the end of Sec.\ref{Str} is very useful for doing this. Using the mapping based on Eqs.(\ref{Effective_width}-\ref{Host_width}) results presented in Fig.\ref{JGNRvsGNR} a)-d) were obtained. For the sake of completeness ribbons Z60 and A120 with and without host ribbons were considered. The presence of the host ribbon in JGNR is our criterion for large values of jag arm indexes $\ell_1$, $\ell_2$ with respect to width index $w$. As can be seen from Fig.\ref{JGNRvsGNR} a) the dependence of $E_g$ on $\varepsilon_x$ for Z60 $\langle 3,3;9 \rangle$ lays between curves for the straight and the effective ribbons. There is no likeliness with the arm ribbon that allows one to obtain high value for $E_g$ of about $0.8$ eV but requires close to SiO$_2$ breakdown electric strength magnitude. Comparing curves for the host and Z60 ribbons, one can see that the former lays below the latter for low values of $\varepsilon_x \sim 0.025$~V/$\mathring{\mbox{A}}$ and above for greater values of $\varepsilon_x$. In general, the curve for Z60 is to a great extend similar to curves for effective, straight and host ribbons especially in region of low values of $\varepsilon_x < 0.02$~V/$\mathring{\mbox{A}}$. However, Fig.\ref{JGNRvsGNR} b) provides quite different results. The curve describing $E_g$ vs. $\varepsilon_x$ dependence for Z60 $\langle 12,12;6 \rangle$ almost coincides with the curve for the arm ribbon ZGNR($6$) for $\varepsilon_x< 0.02$~V/$\mathring{\mbox{A}}$ and considerably deviate from it at higher $\varepsilon_x$. It does not approach curves for the effective ZGNR($24$) or straight ZGNR($12$).
The case of JGNR A120 $\langle 3,3;6 \rangle$ shown in Fig.\ref{JGNRvsGNR} c) significantly differs from that one of JGNRs Z60, because A120 ribbons have armchair edges while they are mapped onto ZGNRs. However, in spite of the fact that the arm ribbon for A120 ribbons has the same type of edges, one notice a crucial difference between them. Firstly, it is impossible to control band gap of AGNR(18), which is the arm one for the A120 $\langle 3,3;6 \rangle$, by means of a transverse electric field. And secondly, the value of the band gap is greater for a jagged ribbon. In the Fig.\ref{JGNRvsGNR} d) one can see the effect of deeper jags which leads to the absence of the host ribbon in A120 $\langle 8,8;4 \rangle$. This JGNR is a bit narrower than that presented in the Fig.\ref{JGNRvsGNR} c), but its band gap is closed nearly at the same magnitude of $\varepsilon_x \sim 0.08$~V/$\mathring{\mbox{A}}$ as for A120 $\langle 3,3;6 \rangle$. It is a straightforward result of jag arm elongation that shifts this point of closing back to lower values of $\varepsilon_x$ after its shift to higher ones due to decrementing of width index $w$. It is also useful to compare curves for A120 ribbons in Fig.\ref{JGNRvsGNR} c) and d) in the region close to zero magnitude of electric field. One sees that while the initial value of the band gap is greater for A120 $\langle 8,8;4 \rangle$, $E_g \sim 0.8$ eV compared to $E_g \sim 0.65$ eV, the closing point is almost the same. Moreover, the drop in the band gap of A120 $\langle 8,8;4 \rangle$ in the region of $\varepsilon_x< 0.02$~V/$\mathring{\mbox{A}}$ is greater than for A120 $\langle 3,3;6 \rangle$ that means the band gap of the former can be more readily controlled. In all other respects Fig.\ref{JGNRvsGNR} c), d) demonstrate very similar curves behaviour, where observable distinctions can be attributed only to quantitative differences of jagged ribbon parameters and, consequently, the straight, effective and arm ribbons ones. For instance, curve for the arm ribbon in Fig.\ref{JGNRvsGNR} d) is also field independent while corresponds to a higher value of the band gap of about $0.6$ eV compared to $0.4$ eV in Fig.\ref{JGNRvsGNR} c).
\begin{figure*}
\begin{center}
\includegraphics[scale=0.7]{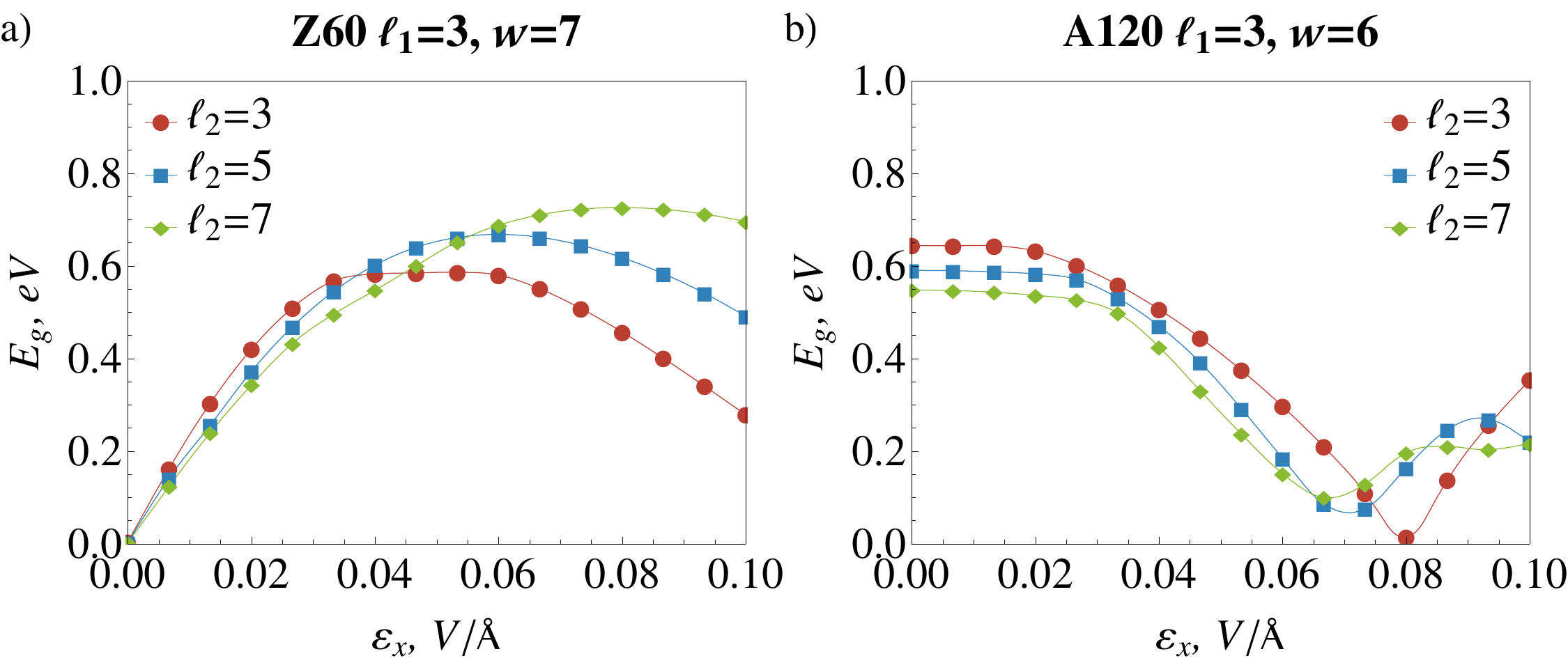}
\end{center}
\caption{\label{Asym}The influence of asymmetry $\ell_1 \neq \ell_2$ on the electric field effect in JGNRs Z60 a) and A120~b).}
\end{figure*}
Finally, we report on the effect of asymmetry $\ell_1 \neq \ell_2$ on both JGNRs Z60 and A120. Having analysed results for JGNRs and ordinary straight GNRs, it is easy to comprehend data shown in Fig.\ref{Asym} a) and b). The increment of just one index $\ell_2$ results in $E_g$ vs. $\varepsilon_x$ curve evolution to the form inherent for the arm ZGNR and AGNR, respectively. To be more persuasive, we suggest the reader to look again at the curves for ZGNR(14) and AGNR(18) in Fig.\ref{JGNRvsGNR} a) and c), correspondingly. After that results in Fig.\ref{Asym} are easy to understand. As in the case of Fig. \ref{BandgapVsElectricfield} for ribbons in Fig.\ref{Asym} we specified linear and quadratic regions of $E_g(\varepsilon_x)$ and calculated parameters (see Table \ref{tab2fig11}). In contrast to the data in Table \ref{tab2fig9}, absolute values of $\beta$ for different ribbons in Table \ref{tab2fig11} are very close and imply approximately equal efficiency in the band gap control.
\begin{table*}
\caption{Fitting of linear and quadratic regions of $E_g(\varepsilon_x)$ curves for JGNRs presented in Fig.\ref{Asym}. Fitting for ribbons Z60$\langle 3,3;7 \rangle$, A120$\langle 3,3;6 \rangle$ is presented in Table \ref{tab2fig9}.}\label{tab2fig11}
\begin{ruledtabular}
\begin{tabular}{cccccccccc}
 & \multicolumn{4}{c}{$\beta x + \alpha $} & \multicolumn{5}{c}{ $\gamma x^2+\beta x+ \alpha$} \\ 
 JGNR & $\alpha$, eV  & $\beta$, eV$\cdot$\AA/V  & $\varepsilon _{\min }$,  V/\AA & $\varepsilon
   _{\max }$,  V/\AA & $\alpha $, eV & $\beta $,  eV$\cdot$\AA/V & $\gamma $, eV$\cdot$\AA$^2$/V$^{2}$ & $ \varepsilon _{\min }$, V/\AA & $\varepsilon _{\max }$, V/\AA \\ \hline
 Z60 $\langle 3,5;7 \rangle$ & 0 & 18.6 & 0 & 0.015 & 0 & 24.1 & -208.9 & 0.015 & 0.06 \\
 Z60 $\langle 3,7;7 \rangle$ & 0 & 18.0 & 0 & 0.010 & 0 & 17.9 & -114.5 & 0.010 & 0.08 \\
 A120 $\langle 3,5;6 \rangle$ & 1.1 & -16.1 & 0.05 & 0.060 & 0.5 & 8.7 & -239.4 & 0.020 & 0.05 \\
 A120 $\langle 3,7;6 \rangle$ & 1.0 & -14.1 & 0.04 & 0.055 & 0.2 & 24.4 & -481.0 & 0.025 & 0.04 \\
\end{tabular}
\end{ruledtabular}
\end{table*}

\section{\label{Conc} Conclusions}
To summarize, in this paper we have studied a subclass of chevron-type GNRs -- jagged graphene nanoribbons (JGNRs). It was shown that it is possible to control the band gap of two types of JGNRs, namely Z60 and A120, by an external transverse electric field, e.g. applied in the plane of the ribbon normally to its longitudinal axis. The band gap opening is possible for Z60 ribbons, while the band gap closing is characteristic for A120 ribbons. In both cases the value of the band gap opening/closing is greater for narrower ribbons and requires stronger electric fields. As there is a natural limit for electric strength values due to the breakdown phenomenon, there must be an optimal value of ribbon width providing the highest possible value of the band gap. We argue that for $\varepsilon_{x,max}=0.1$~V/$\mathring{\mbox{A}}$, e.g. SiO$_2$ as a substrate, the optimal width is about $w \sim 6$. For these values of width Z60 ribbons in a transverse electric field behave like a quantum dot system and series of emission and absorption lines must be observed if they are not forbidden by optical selection rules. Some enhancement in the controllability of the band gap in the transverse electric field is achievable for A120 ribbons as the longer jag arms the lower field is required to close the band gap, while it is not the case for Z60 ones. Also we must mentioned the second peak for A120 JGNR that is seemed to be unique for this type of structures as it was not mentioned for ordinary straight GNRs in the paper\cite{Chang2006}.
Finally, we notice that the value of band gap opening for Z60 is about $0.6$ eV and so high that even taking into account possible error in band gap evaluation ($\approx 10$\%) and assuming the same field screening ($\sim 25$\% for $\varepsilon_x = 0.05 $ V/$\mathring{\mbox{A}}$) as for bilayer graphene one can assess band gap opening as no less than $\sim 0.4$ eV that is about twice larger than for bilayer graphene in a normal electric field\cite{Castro2007}. The value is also large compared to that one for carbon nanotubes and can be implemented for fabrication of all metallic transistors proposed firstly for nanotubes in the paper\cite{Rotkin2004}. Following the just mentioned work, it seems reasonable even to combine normal and transverse geometry to control conductivity of the channel and carrier concentration separately. Actually, separate control of the band gap and the Fermi energy paves the way to a new type of devices that could build a bridge between optics and electronics. Speaking again of A120, one must stress also the quite wide range of band gap variation of about $0.6$ eV. This property of A120 ribbons could make them suitable candidates for infrared or THz laser media with widely tunable irradiation frequency. Although only a homogeneous transverse electric field was considered we deem that the results obtained give a hint on the effect of charge impurities on some types of GNRs, however to prove the last statement another detailed study is necessary. It is quite important to notice that in spite of the fact that only graphene nanoribbons were considered in this study, its result can be relevant to other structures with similar geometry produced by patterned evaporation\cite{Wei2010} of H, O, F, atoms from the surface of graphane, graphene oxide or graphene fluoride, respectively, or substitution of BN atoms on C atoms in h-BN layer\cite{Liu2013a}. As was shown in papers \cite{Chernozatonskii2007,Chernozatonskii2010,Ribas2010}, such structures exhibit similar electronic properties to GNRs of the same shape. After having written this paper we became aware of recent results confirming spin ordering on the zigzag edges at room temperature\cite{Magda2014}. This effect can reduce the range of tunability for Z60 ribbons and needs a separate detailed investigation. 

\begin{acknowledgments}
This work was supported by EU FP7 ITN NOTEDEV (through Grant No. FP7-607521); IRSES projects CACOMEL (Grant No. FP7-247007), FAEMCAR (Grant No. FP7-318617) and CANTOR (Grant No. FP7-612285); Graphene Flagship (Grant No. 604391) and the Ministry of Education of the Republic of Belarus (Grant No.  20140773). The authors are very grateful to Prof. Philippe Lambin and Prof. Mikhail Portnoi for useful advice and Charles Downing for careful reading of the manuscript .
\end{acknowledgments}

\appendix*
\section{\label{App} Realistic Z60 structure}
\begin{figure*}
\begin{center}
\includegraphics[scale=0.7]{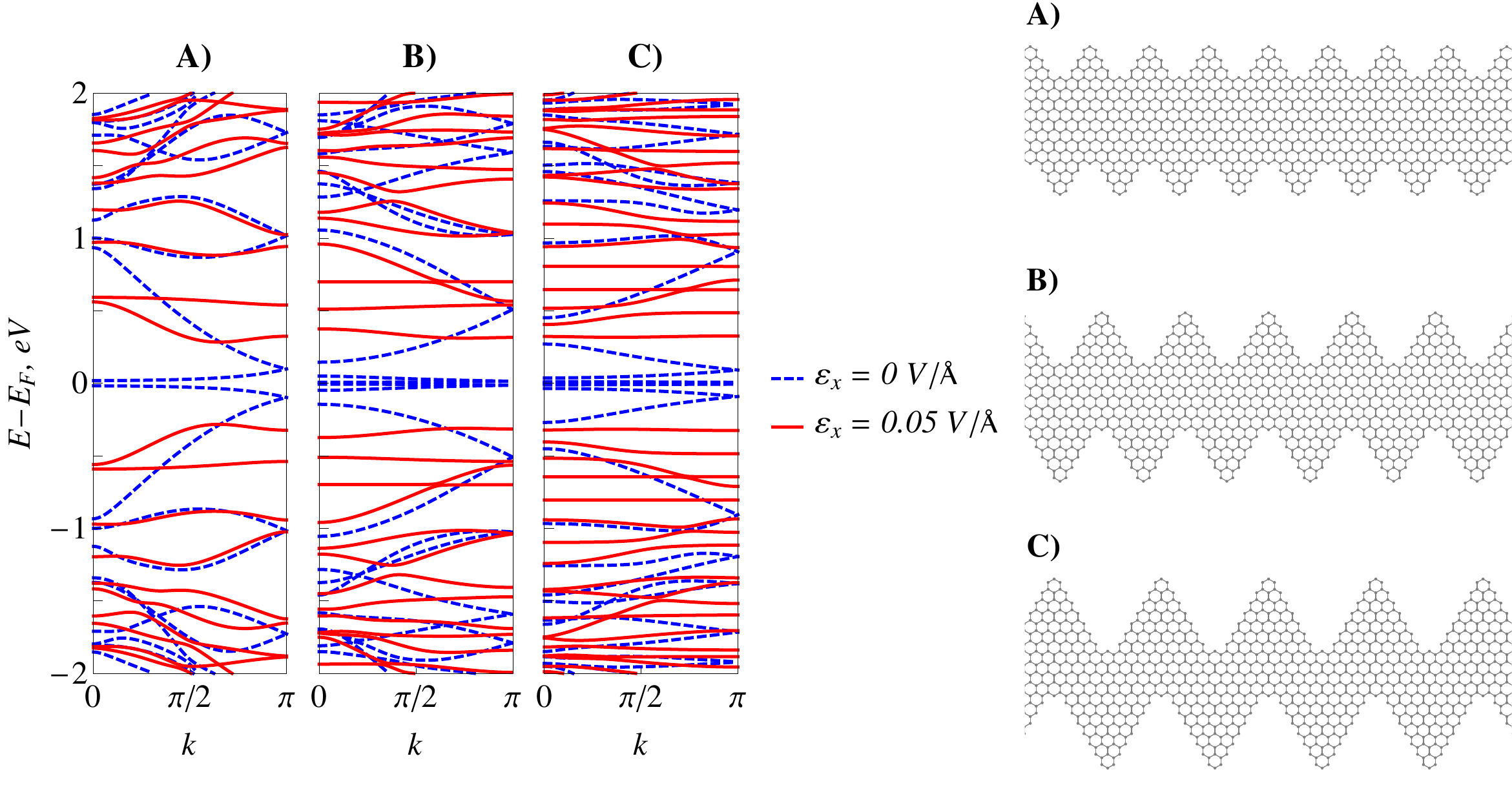}
\end{center}
\caption{\label{SplittingZ60r}The splitting of dispersionless bands in the transverse electric field for JGNRs Z60r.}
\end{figure*}

Ribbons of Z60 type are good as a theoretical model but unfortunately it is very likely that such configurations are difficult to realize experimentally. Therefore we decided to provide additional results to prove that the found effect is not due to the peculiar structural configuration associated with one carbon atom missing in hexagons on the border between two jags but rather due the ribbon chevron-type pattern and zigzag edge. A structure with one carbon atom removed from the jag apex and placed at the opposite edge to complete hexagons seems to be more stable. Such structure can be referred to as Z60r, where "r" -- stand for "realistic". As one can see in Fig. \ref{SplittingZ60r}, band gap opening and dispersionless band splitting in the external transverse electric field can be found as well in structures of more energetically favourable configuration.

\bibliography{library}

\begin{thebibliography}{62}%
\makeatletter
\providecommand \@ifxundefined [1]{%
 \@ifx{#1\undefined}
}%
\providecommand \@ifnum [1]{%
 \ifnum #1\expandafter \@firstoftwo
 \else \expandafter \@secondoftwo
 \fi
}%
\providecommand \@ifx [1]{%
 \ifx #1\expandafter \@firstoftwo
 \else \expandafter \@secondoftwo
 \fi
}%
\providecommand \natexlab [1]{#1}%
\providecommand \enquote  [1]{``#1''}%
\providecommand \bibnamefont  [1]{#1}%
\providecommand \bibfnamefont [1]{#1}%
\providecommand \citenamefont [1]{#1}%
\providecommand \href@noop [0]{\@secondoftwo}%
\providecommand \href [0]{\begingroup \@sanitize@url \@href}%
\providecommand \@href[1]{\@@startlink{#1}\@@href}%
\providecommand \@@href[1]{\endgroup#1\@@endlink}%
\providecommand \@sanitize@url [0]{\catcode `\\12\catcode `\$12\catcode
  `\&12\catcode `\#12\catcode `\^12\catcode `\_12\catcode `\%12\relax}%
\providecommand \@@startlink[1]{}%
\providecommand \@@endlink[0]{}%
\providecommand \url  [0]{\begingroup\@sanitize@url \@url }%
\providecommand \@url [1]{\endgroup\@href {#1}{\urlprefix }}%
\providecommand \urlprefix  [0]{URL }%
\providecommand \Eprint [0]{\href }%
\providecommand \doibase [0]{http://dx.doi.org/}%
\providecommand \selectlanguage [0]{\@gobble}%
\providecommand \bibinfo  [0]{\@secondoftwo}%
\providecommand \bibfield  [0]{\@secondoftwo}%
\providecommand \translation [1]{[#1]}%
\providecommand \BibitemOpen [0]{}%
\providecommand \bibitemStop [0]{}%
\providecommand \bibitemNoStop [0]{.\EOS\space}%
\providecommand \EOS [0]{\spacefactor3000\relax}%
\providecommand \BibitemShut  [1]{\csname bibitem#1\endcsname}%
\let\auto@bib@innerbib\@empty
\bibitem [{\citenamefont {Novoselov}\ \emph {et~al.}(2004)\citenamefont
  {Novoselov}, \citenamefont {Geim}, \citenamefont {Morozov}, \citenamefont
  {Jiang}, \citenamefont {Zhang}, \citenamefont {Dubonos}, \citenamefont
  {Grigorieva},\ and\ \citenamefont {Firsov}}]{Novoselov2004}%
  \BibitemOpen
  \bibfield  {author} {\bibinfo {author} {\bibfnamefont {K.~S.}\ \bibnamefont
  {Novoselov}}, \bibinfo {author} {\bibfnamefont {A.~K.}\ \bibnamefont {Geim}},
  \bibinfo {author} {\bibfnamefont {S.~V.}\ \bibnamefont {Morozov}}, \bibinfo
  {author} {\bibfnamefont {D.}~\bibnamefont {Jiang}}, \bibinfo {author}
  {\bibfnamefont {Y.}~\bibnamefont {Zhang}}, \bibinfo {author} {\bibfnamefont
  {S.~V.}\ \bibnamefont {Dubonos}}, \bibinfo {author} {\bibfnamefont {I.~V.}\
  \bibnamefont {Grigorieva}}, \ and\ \bibinfo {author} {\bibfnamefont {A.~A.}\
  \bibnamefont {Firsov}},\ }\href {\doibase 10.1126/science.1102896} {\bibfield
   {journal} {\bibinfo  {journal} {Science (New York, N.Y.)}\ }\textbf
  {\bibinfo {volume} {306}},\ \bibinfo {pages} {666} (\bibinfo {year}
  {2004})}\BibitemShut {NoStop}%
\bibitem [{\citenamefont {Novoselov}\ \emph {et~al.}(2005)\citenamefont
  {Novoselov}, \citenamefont {Geim}, \citenamefont {Morozov}, \citenamefont
  {Jiang}, \citenamefont {Katsnelson}, \citenamefont {Grigorieva},
  \citenamefont {Dubonos},\ and\ \citenamefont {Firsov}}]{Novoselov2005}%
  \BibitemOpen
  \bibfield  {author} {\bibinfo {author} {\bibfnamefont {K.~S.}\ \bibnamefont
  {Novoselov}}, \bibinfo {author} {\bibfnamefont {A.~K.}\ \bibnamefont {Geim}},
  \bibinfo {author} {\bibfnamefont {S.~V.}\ \bibnamefont {Morozov}}, \bibinfo
  {author} {\bibfnamefont {D.}~\bibnamefont {Jiang}}, \bibinfo {author}
  {\bibfnamefont {M.~I.}\ \bibnamefont {Katsnelson}}, \bibinfo {author}
  {\bibfnamefont {I.~V.}\ \bibnamefont {Grigorieva}}, \bibinfo {author}
  {\bibfnamefont {S.~V.}\ \bibnamefont {Dubonos}}, \ and\ \bibinfo {author}
  {\bibfnamefont {A.~A.}\ \bibnamefont {Firsov}},\ }\href {\doibase
  10.1038/nature04233} {\bibfield  {journal} {\bibinfo  {journal} {Nature}\
  }\textbf {\bibinfo {volume} {438}},\ \bibinfo {pages} {197} (\bibinfo {year}
  {2005})}\BibitemShut {NoStop}%
\bibitem [{\citenamefont {Lee}\ \emph {et~al.}(2012)\citenamefont {Lee},
  \citenamefont {Yoon},\ and\ \citenamefont {Cheong}}]{Lee2012}%
  \BibitemOpen
  \bibfield  {author} {\bibinfo {author} {\bibfnamefont {J.-U.}\ \bibnamefont
  {Lee}}, \bibinfo {author} {\bibfnamefont {D.}~\bibnamefont {Yoon}}, \ and\
  \bibinfo {author} {\bibfnamefont {H.}~\bibnamefont {Cheong}},\ }\href
  {\doibase 10.1021/nl301073q} {\bibfield  {journal} {\bibinfo  {journal} {Nano
  letters}\ }\textbf {\bibinfo {volume} {12}},\ \bibinfo {pages} {4444}
  (\bibinfo {year} {2012})}\BibitemShut {NoStop}%
\bibitem [{\citenamefont {Balandin}\ \emph {et~al.}(2008)\citenamefont
  {Balandin}, \citenamefont {Ghosh}, \citenamefont {Bao}, \citenamefont
  {Calizo}, \citenamefont {Teweldebrhan}, \citenamefont {Miao},\ and\
  \citenamefont {Lau}}]{Balandin2008}%
  \BibitemOpen
  \bibfield  {author} {\bibinfo {author} {\bibfnamefont {A.~A.}\ \bibnamefont
  {Balandin}}, \bibinfo {author} {\bibfnamefont {S.}~\bibnamefont {Ghosh}},
  \bibinfo {author} {\bibfnamefont {W.}~\bibnamefont {Bao}}, \bibinfo {author}
  {\bibfnamefont {I.}~\bibnamefont {Calizo}}, \bibinfo {author} {\bibfnamefont
  {D.}~\bibnamefont {Teweldebrhan}}, \bibinfo {author} {\bibfnamefont
  {F.}~\bibnamefont {Miao}}, \ and\ \bibinfo {author} {\bibfnamefont {C.~N.}\
  \bibnamefont {Lau}},\ }\href {\doibase 10.1021/nl0731872} {\bibfield
  {journal} {\bibinfo  {journal} {Nano letters}\ }\textbf {\bibinfo {volume}
  {8}},\ \bibinfo {pages} {902} (\bibinfo {year} {2008})}\BibitemShut {NoStop}%
\bibitem [{\citenamefont {Chen}\ \emph {et~al.}(2010)\citenamefont {Chen},
  \citenamefont {Jayasekera}, \citenamefont {Calzolari}, \citenamefont {Kim},\
  and\ \citenamefont {Nardelli}}]{Chen2010}%
  \BibitemOpen
  \bibfield  {author} {\bibinfo {author} {\bibfnamefont {Y.}~\bibnamefont
  {Chen}}, \bibinfo {author} {\bibfnamefont {T.}~\bibnamefont {Jayasekera}},
  \bibinfo {author} {\bibfnamefont {A.}~\bibnamefont {Calzolari}}, \bibinfo
  {author} {\bibfnamefont {K.~W.}\ \bibnamefont {Kim}}, \ and\ \bibinfo
  {author} {\bibfnamefont {M.~B.}\ \bibnamefont {Nardelli}},\ }\href {\doibase
  10.1088/0953-8984/22/37/372202} {\bibfield  {journal} {\bibinfo  {journal}
  {Journal of physics. Condensed matter : an Institute of Physics journal}\
  }\textbf {\bibinfo {volume} {22}},\ \bibinfo {pages} {372202} (\bibinfo
  {year} {2010})}\BibitemShut {NoStop}%
\bibitem [{\citenamefont {Chernozatonskii}\ \emph {et~al.}(2014)\citenamefont
  {Chernozatonskii}, \citenamefont {Sorokin},\ and\ \citenamefont
  {Artukh}}]{Chernozatonskii2014}%
  \BibitemOpen
  \bibfield  {author} {\bibinfo {author} {\bibfnamefont {L.~A.}\ \bibnamefont
  {Chernozatonskii}}, \bibinfo {author} {\bibfnamefont {P.~B.}\ \bibnamefont
  {Sorokin}}, \ and\ \bibinfo {author} {\bibfnamefont {A.~A.}\ \bibnamefont
  {Artukh}},\ }\href {\doibase 10.1070/RC2014v083n03ABEH004367} {\bibfield
  {journal} {\bibinfo  {journal} {Russian Chemical Reviews}\ }\textbf {\bibinfo
  {volume} {83}},\ \bibinfo {pages} {251} (\bibinfo {year} {2014})}\BibitemShut
  {NoStop}%
\bibitem [{\citenamefont {Schwierz}(2010)}]{Schwierz2010}%
  \BibitemOpen
  \bibfield  {author} {\bibinfo {author} {\bibfnamefont {F.}~\bibnamefont
  {Schwierz}},\ }\href {\doibase 10.1038/nnano.2010.89} {\bibfield  {journal}
  {\bibinfo  {journal} {Nature nanotechnology}\ }\textbf {\bibinfo {volume}
  {5}},\ \bibinfo {pages} {487} (\bibinfo {year} {2010})}\BibitemShut {NoStop}%
\bibitem [{\citenamefont {Murali}\ \emph {et~al.}(2009)\citenamefont {Murali},
  \citenamefont {Yang}, \citenamefont {Brenner}, \citenamefont {Beck},\ and\
  \citenamefont {Meindl}}]{Murali2009a}%
  \BibitemOpen
  \bibfield  {author} {\bibinfo {author} {\bibfnamefont {R.}~\bibnamefont
  {Murali}}, \bibinfo {author} {\bibfnamefont {Y.}~\bibnamefont {Yang}},
  \bibinfo {author} {\bibfnamefont {K.}~\bibnamefont {Brenner}}, \bibinfo
  {author} {\bibfnamefont {T.}~\bibnamefont {Beck}}, \ and\ \bibinfo {author}
  {\bibfnamefont {J.~D.}\ \bibnamefont {Meindl}},\ }\href {\doibase
  10.1063/1.3147183} {\bibfield  {journal} {\bibinfo  {journal} {Applied
  Physics Letters}\ }\textbf {\bibinfo {volume} {94}},\ \bibinfo {pages}
  {243114} (\bibinfo {year} {2009})}\BibitemShut {NoStop}%
\bibitem [{\citenamefont {Maffucci}\ and\ \citenamefont
  {Miano}(2013)}]{Interconnects2013}%
  \BibitemOpen
  \bibfield  {author} {\bibinfo {author} {\bibfnamefont {A.}~\bibnamefont
  {Maffucci}}\ and\ \bibinfo {author} {\bibfnamefont {G.}~\bibnamefont
  {Miano}},\ }\href {\doibase 10.1109/TNANO.2013.2274901} {\bibfield  {journal}
  {\bibinfo  {journal} {IEEE Transactions on Nanotechnology}\ }\textbf
  {\bibinfo {volume} {12}},\ \bibinfo {pages} {817} (\bibinfo {year}
  {2013})}\BibitemShut {NoStop}%
\bibitem [{\citenamefont {Schedin}\ \emph {et~al.}(2007)\citenamefont
  {Schedin}, \citenamefont {Geim}, \citenamefont {Morozov}, \citenamefont
  {Hill}, \citenamefont {Blake}, \citenamefont {Katsnelson},\ and\
  \citenamefont {Novoselov}}]{Schedin2007}%
  \BibitemOpen
  \bibfield  {author} {\bibinfo {author} {\bibfnamefont {F.}~\bibnamefont
  {Schedin}}, \bibinfo {author} {\bibfnamefont {A.~K.}\ \bibnamefont {Geim}},
  \bibinfo {author} {\bibfnamefont {S.~V.}\ \bibnamefont {Morozov}}, \bibinfo
  {author} {\bibfnamefont {E.~W.}\ \bibnamefont {Hill}}, \bibinfo {author}
  {\bibfnamefont {P.}~\bibnamefont {Blake}}, \bibinfo {author} {\bibfnamefont
  {M.~I.}\ \bibnamefont {Katsnelson}}, \ and\ \bibinfo {author} {\bibfnamefont
  {K.~S.}\ \bibnamefont {Novoselov}},\ }\href {\doibase 10.1038/nmat1967}
  {\bibfield  {journal} {\bibinfo  {journal} {Nature materials}\ }\textbf
  {\bibinfo {volume} {6}},\ \bibinfo {pages} {652} (\bibinfo {year}
  {2007})}\BibitemShut {NoStop}%
\bibitem [{\citenamefont {Bi}\ \emph {et~al.}(2013)\citenamefont {Bi},
  \citenamefont {Yin}, \citenamefont {Xie}, \citenamefont {Ji}, \citenamefont
  {Wan}, \citenamefont {Sun}, \citenamefont {Terrones},\ and\ \citenamefont
  {Dresselhaus}}]{Bi2013}%
  \BibitemOpen
  \bibfield  {author} {\bibinfo {author} {\bibfnamefont {H.}~\bibnamefont
  {Bi}}, \bibinfo {author} {\bibfnamefont {K.}~\bibnamefont {Yin}}, \bibinfo
  {author} {\bibfnamefont {X.}~\bibnamefont {Xie}}, \bibinfo {author}
  {\bibfnamefont {J.}~\bibnamefont {Ji}}, \bibinfo {author} {\bibfnamefont
  {S.}~\bibnamefont {Wan}}, \bibinfo {author} {\bibfnamefont {L.}~\bibnamefont
  {Sun}}, \bibinfo {author} {\bibfnamefont {M.}~\bibnamefont {Terrones}}, \
  and\ \bibinfo {author} {\bibfnamefont {M.~S.}\ \bibnamefont {Dresselhaus}},\
  }\href {\doibase 10.1038/srep02714} {\bibfield  {journal} {\bibinfo
  {journal} {Scientific reports}\ }\textbf {\bibinfo {volume} {3}},\ \bibinfo
  {pages} {2714} (\bibinfo {year} {2013})}\BibitemShut {NoStop}%
\bibitem [{\citenamefont {Tozzini}\ and\ \citenamefont
  {Pellegrini}(2013)}]{Tozzini2013}%
  \BibitemOpen
  \bibfield  {author} {\bibinfo {author} {\bibfnamefont {V.}~\bibnamefont
  {Tozzini}}\ and\ \bibinfo {author} {\bibfnamefont {V.}~\bibnamefont
  {Pellegrini}},\ }\href {\doibase 10.1039/c2cp42538f} {\bibfield  {journal}
  {\bibinfo  {journal} {Physical chemistry chemical physics : PCCP}\ }\textbf
  {\bibinfo {volume} {15}},\ \bibinfo {pages} {80} (\bibinfo {year}
  {2013})}\BibitemShut {NoStop}%
\bibitem [{\citenamefont {Mikhailov}(2013)}]{Mikhailov2012}%
  \BibitemOpen
  \bibfield  {author} {\bibinfo {author} {\bibfnamefont {S.}~\bibnamefont
  {Mikhailov}},\ }\href {\doibase 10.1103/PhysRevB.87.115405} {\bibfield
  {journal} {\bibinfo  {journal} {Physical Review B}\ }\textbf {\bibinfo
  {volume} {87}},\ \bibinfo {pages} {115405} (\bibinfo {year}
  {2013})}\BibitemShut {NoStop}%
\bibitem [{\citenamefont {Batrakov}\ \emph {et~al.}(2012)\citenamefont
  {Batrakov}, \citenamefont {Saroka}, \citenamefont {Maksimenko},\ and\
  \citenamefont {Thomsen}}]{Batrakov2012}%
  \BibitemOpen
  \bibfield  {author} {\bibinfo {author} {\bibfnamefont {K.~G.}\ \bibnamefont
  {Batrakov}}, \bibinfo {author} {\bibfnamefont {V.~A.}\ \bibnamefont
  {Saroka}}, \bibinfo {author} {\bibfnamefont {S.~A.}\ \bibnamefont
  {Maksimenko}}, \ and\ \bibinfo {author} {\bibfnamefont {C.}~\bibnamefont
  {Thomsen}},\ }\href {\doibase 10.1117/1.JNP.6.061719} {\bibfield  {journal}
  {\bibinfo  {journal} {Journal of Nanophotonics}\ }\textbf {\bibinfo {volume}
  {6}},\ \bibinfo {pages} {061719} (\bibinfo {year} {2012})}\BibitemShut
  {NoStop}%
\bibitem [{\citenamefont {Hartmann}\ \emph {et~al.}(2014)\citenamefont
  {Hartmann}, \citenamefont {Kono},\ and\ \citenamefont
  {Portnoi}}]{Hartmann2014}%
  \BibitemOpen
  \bibfield  {author} {\bibinfo {author} {\bibfnamefont {R.~R.}\ \bibnamefont
  {Hartmann}}, \bibinfo {author} {\bibfnamefont {J.}~\bibnamefont {Kono}}, \
  and\ \bibinfo {author} {\bibfnamefont {M.~E.}\ \bibnamefont {Portnoi}},\
  }\href {\doibase 10.1088/0957-4484/25/32/322001} {\bibfield  {journal}
  {\bibinfo  {journal} {Nanotechnology}\ }\textbf {\bibinfo {volume} {25}},\
  \bibinfo {pages} {322001} (\bibinfo {year} {2014})}\BibitemShut {NoStop}%
\bibitem [{\citenamefont {Smith}\ \emph {et~al.}(2013)\citenamefont {Smith},
  \citenamefont {Franklin}, \citenamefont {Farmer},\ and\ \citenamefont
  {Dimitrakopoulos}}]{Smith2013}%
  \BibitemOpen
  \bibfield  {author} {\bibinfo {author} {\bibfnamefont {J.~T.}\ \bibnamefont
  {Smith}}, \bibinfo {author} {\bibfnamefont {A.~D.}\ \bibnamefont {Franklin}},
  \bibinfo {author} {\bibfnamefont {D.~B.}\ \bibnamefont {Farmer}}, \ and\
  \bibinfo {author} {\bibfnamefont {C.~D.}\ \bibnamefont {Dimitrakopoulos}},\
  }\href {\doibase 10.1021/nn400671z} {\bibfield  {journal} {\bibinfo
  {journal} {ACS nano}\ }\textbf {\bibinfo {volume} {7}},\ \bibinfo {pages}
  {3661} (\bibinfo {year} {2013})}\BibitemShut {NoStop}%
\bibitem [{\citenamefont {Liang}\ \emph {et~al.}(2010)\citenamefont {Liang},
  \citenamefont {Chen}, \citenamefont {Xu}, \citenamefont {Liu}, \citenamefont
  {Zhang}, \citenamefont {Zhao}, \citenamefont {Zhang}, \citenamefont {Tian},
  \citenamefont {Huang}, \citenamefont {Ma},\ and\ \citenamefont
  {Li}}]{Liang2010}%
  \BibitemOpen
  \bibfield  {author} {\bibinfo {author} {\bibfnamefont {J.}~\bibnamefont
  {Liang}}, \bibinfo {author} {\bibfnamefont {Y.}~\bibnamefont {Chen}},
  \bibinfo {author} {\bibfnamefont {Y.}~\bibnamefont {Xu}}, \bibinfo {author}
  {\bibfnamefont {Z.}~\bibnamefont {Liu}}, \bibinfo {author} {\bibfnamefont
  {L.}~\bibnamefont {Zhang}}, \bibinfo {author} {\bibfnamefont
  {X.}~\bibnamefont {Zhao}}, \bibinfo {author} {\bibfnamefont {X.}~\bibnamefont
  {Zhang}}, \bibinfo {author} {\bibfnamefont {J.}~\bibnamefont {Tian}},
  \bibinfo {author} {\bibfnamefont {Y.}~\bibnamefont {Huang}}, \bibinfo
  {author} {\bibfnamefont {Y.}~\bibnamefont {Ma}}, \ and\ \bibinfo {author}
  {\bibfnamefont {F.}~\bibnamefont {Li}},\ }\href {\doibase 10.1021/am1007326}
  {\bibfield  {journal} {\bibinfo  {journal} {ACS applied materials \&
  interfaces}\ }\textbf {\bibinfo {volume} {2}},\ \bibinfo {pages} {3310}
  (\bibinfo {year} {2010})}\BibitemShut {NoStop}%
\bibitem [{\citenamefont {Dvorak}\ \emph {et~al.}(2013)\citenamefont {Dvorak},
  \citenamefont {Oswald},\ and\ \citenamefont {Wu}}]{Dvorak2013}%
  \BibitemOpen
  \bibfield  {author} {\bibinfo {author} {\bibfnamefont {M.}~\bibnamefont
  {Dvorak}}, \bibinfo {author} {\bibfnamefont {W.}~\bibnamefont {Oswald}}, \
  and\ \bibinfo {author} {\bibfnamefont {Z.}~\bibnamefont {Wu}},\ }\href
  {\doibase 10.1038/srep02289} {\bibfield  {journal} {\bibinfo  {journal}
  {Scientific reports}\ }\textbf {\bibinfo {volume} {3}},\ \bibinfo {pages}
  {2289} (\bibinfo {year} {2013})}\BibitemShut {NoStop}%
\bibitem [{\citenamefont {Chernozatonskii}\ \emph {et~al.}(2007)\citenamefont
  {Chernozatonskii}, \citenamefont {Sorokin},\ and\ \citenamefont
  {Brüning}}]{Chernozatonskii2007}%
  \BibitemOpen
  \bibfield  {author} {\bibinfo {author} {\bibfnamefont {L.~A.}\ \bibnamefont
  {Chernozatonskii}}, \bibinfo {author} {\bibfnamefont {P.~B.}\ \bibnamefont
  {Sorokin}}, \ and\ \bibinfo {author} {\bibfnamefont {J.~W.}\ \bibnamefont
  {Brüning}},\ }\href {\doibase 10.1063/1.2800889} {\bibfield  {journal}
  {\bibinfo  {journal} {Applied Physics Letters}\ }\textbf {\bibinfo {volume}
  {91}},\ \bibinfo {pages} {183103} (\bibinfo {year} {2007})}\BibitemShut
  {NoStop}%
\bibitem [{\citenamefont {Choi}\ \emph {et~al.}(2010)\citenamefont {Choi},
  \citenamefont {Jhi},\ and\ \citenamefont {Son}}]{Choi2010}%
  \BibitemOpen
  \bibfield  {author} {\bibinfo {author} {\bibfnamefont {S.-M.}\ \bibnamefont
  {Choi}}, \bibinfo {author} {\bibfnamefont {S.-H.}\ \bibnamefont {Jhi}}, \
  and\ \bibinfo {author} {\bibfnamefont {Y.-W.}\ \bibnamefont {Son}},\ }\href
  {\doibase 10.1103/PhysRevB.81.081407} {\bibfield  {journal} {\bibinfo
  {journal} {Physical Review B}\ }\textbf {\bibinfo {volume} {81}},\ \bibinfo
  {pages} {081407} (\bibinfo {year} {2010})}\BibitemShut {NoStop}%
\bibitem [{\citenamefont {Li}\ \emph {et~al.}(2010)\citenamefont {Li},
  \citenamefont {Jiang}, \citenamefont {Liu},\ and\ \citenamefont
  {Liu}}]{Li2010}%
  \BibitemOpen
  \bibfield  {author} {\bibinfo {author} {\bibfnamefont {Y.}~\bibnamefont
  {Li}}, \bibinfo {author} {\bibfnamefont {X.}~\bibnamefont {Jiang}}, \bibinfo
  {author} {\bibfnamefont {Z.}~\bibnamefont {Liu}}, \ and\ \bibinfo {author}
  {\bibfnamefont {Z.}~\bibnamefont {Liu}},\ }\href {\doibase
  10.1007/s12274-010-0015-7} {\bibfield  {journal} {\bibinfo  {journal} {Nano
  Research}\ }\textbf {\bibinfo {volume} {3}},\ \bibinfo {pages} {545}
  (\bibinfo {year} {2010})}\BibitemShut {NoStop}%
\bibitem [{\citenamefont {Ribeiro}\ \emph {et~al.}(2009)\citenamefont
  {Ribeiro}, \citenamefont {Pereira}, \citenamefont {Peres}, \citenamefont
  {Briddon},\ and\ \citenamefont {{Castro Neto}}}]{Ribeiro2009}%
  \BibitemOpen
  \bibfield  {author} {\bibinfo {author} {\bibfnamefont {R.~M.}\ \bibnamefont
  {Ribeiro}}, \bibinfo {author} {\bibfnamefont {V.~M.}\ \bibnamefont
  {Pereira}}, \bibinfo {author} {\bibfnamefont {N.~M.~R.}\ \bibnamefont
  {Peres}}, \bibinfo {author} {\bibfnamefont {P.~R.}\ \bibnamefont {Briddon}},
  \ and\ \bibinfo {author} {\bibfnamefont {A.~H.}\ \bibnamefont {{Castro
  Neto}}},\ }\href {\doibase 10.1088/1367-2630/11/11/115002} {\bibfield
  {journal} {\bibinfo  {journal} {New Journal of Physics}\ }\textbf {\bibinfo
  {volume} {11}},\ \bibinfo {pages} {115002} (\bibinfo {year}
  {2009})}\BibitemShut {NoStop}%
\bibitem [{\citenamefont {Pereira}\ and\ \citenamefont {{Castro
  Neto}}(2009)}]{Pereira2009}%
  \BibitemOpen
  \bibfield  {author} {\bibinfo {author} {\bibfnamefont {V.}~\bibnamefont
  {Pereira}}\ and\ \bibinfo {author} {\bibfnamefont {A.}~\bibnamefont {{Castro
  Neto}}},\ }\href {\doibase 10.1103/PhysRevLett.103.046801} {\bibfield
  {journal} {\bibinfo  {journal} {Physical Review Letters}\ }\textbf {\bibinfo
  {volume} {103}},\ \bibinfo {pages} {046801} (\bibinfo {year}
  {2009})}\BibitemShut {NoStop}%
\bibitem [{\citenamefont {Chernozatonskii}\ and\ \citenamefont
  {Sorokin}(2010)}]{Chernozatonskii2010}%
  \BibitemOpen
  \bibfield  {author} {\bibinfo {author} {\bibfnamefont {L.~A.}\ \bibnamefont
  {Chernozatonskii}}\ and\ \bibinfo {author} {\bibfnamefont {P.~B.}\
  \bibnamefont {Sorokin}},\ }\href {\doibase 10.1021/jp9100653} {\bibfield
  {journal} {\bibinfo  {journal} {The Journal of Physical Chemistry C}\
  }\textbf {\bibinfo {volume} {114}},\ \bibinfo {pages} {3225} (\bibinfo {year}
  {2010})}\BibitemShut {NoStop}%
\bibitem [{\citenamefont {McCann}(2006)}]{McCann2006a}%
  \BibitemOpen
  \bibfield  {author} {\bibinfo {author} {\bibfnamefont {E.}~\bibnamefont
  {McCann}},\ }\href {\doibase 10.1103/PhysRevB.74.161403} {\bibfield
  {journal} {\bibinfo  {journal} {Physical Review B}\ }\textbf {\bibinfo
  {volume} {74}},\ \bibinfo {pages} {161403} (\bibinfo {year}
  {2006})}\BibitemShut {NoStop}%
\bibitem [{\citenamefont {Castro}\ \emph {et~al.}(2007)\citenamefont {Castro},
  \citenamefont {Novoselov}, \citenamefont {Morozov}, \citenamefont {Peres},
  \citenamefont {dos Santos}, \citenamefont {Nilsson}, \citenamefont {Guinea},
  \citenamefont {Geim},\ and\ \citenamefont {Neto}}]{Castro2007}%
  \BibitemOpen
  \bibfield  {author} {\bibinfo {author} {\bibfnamefont {E.}~\bibnamefont
  {Castro}}, \bibinfo {author} {\bibfnamefont {K.}~\bibnamefont {Novoselov}},
  \bibinfo {author} {\bibfnamefont {S.}~\bibnamefont {Morozov}}, \bibinfo
  {author} {\bibfnamefont {N.}~\bibnamefont {Peres}}, \bibinfo {author}
  {\bibfnamefont {J.}~\bibnamefont {dos Santos}}, \bibinfo {author}
  {\bibfnamefont {J.}~\bibnamefont {Nilsson}}, \bibinfo {author} {\bibfnamefont
  {F.}~\bibnamefont {Guinea}}, \bibinfo {author} {\bibfnamefont
  {A.}~\bibnamefont {Geim}}, \ and\ \bibinfo {author} {\bibfnamefont
  {A.}~\bibnamefont {Neto}},\ }\href {\doibase 10.1103/PhysRevLett.99.216802}
  {\bibfield  {journal} {\bibinfo  {journal} {Physical Review Letters}\
  }\textbf {\bibinfo {volume} {99}},\ \bibinfo {pages} {216802} (\bibinfo
  {year} {2007})}\BibitemShut {NoStop}%
\bibitem [{\citenamefont {Lui}\ \emph {et~al.}(2011)\citenamefont {Lui},
  \citenamefont {Li}, \citenamefont {Mak}, \citenamefont {Cappelluti},\ and\
  \citenamefont {Heinz}}]{Lui}%
  \BibitemOpen
  \bibfield  {author} {\bibinfo {author} {\bibfnamefont {C.~H.}\ \bibnamefont
  {Lui}}, \bibinfo {author} {\bibfnamefont {Z.}~\bibnamefont {Li}}, \bibinfo
  {author} {\bibfnamefont {K.~F.}\ \bibnamefont {Mak}}, \bibinfo {author}
  {\bibfnamefont {E.}~\bibnamefont {Cappelluti}}, \ and\ \bibinfo {author}
  {\bibfnamefont {T.~F.}\ \bibnamefont {Heinz}},\ }\href {\doibase
  10.1038/nphys2102} {\bibfield  {journal} {\bibinfo  {journal} {Nature
  Physics}\ }\textbf {\bibinfo {volume} {7}},\ \bibinfo {pages} {944} (\bibinfo
  {year} {2011})}\BibitemShut {NoStop}%
\bibitem [{\citenamefont {Sahu}\ \emph {et~al.}(2010)\citenamefont {Sahu},
  \citenamefont {Min},\ and\ \citenamefont {Banerjee}}]{Sahu2010a}%
  \BibitemOpen
  \bibfield  {author} {\bibinfo {author} {\bibfnamefont {B.}~\bibnamefont
  {Sahu}}, \bibinfo {author} {\bibfnamefont {H.}~\bibnamefont {Min}}, \ and\
  \bibinfo {author} {\bibfnamefont {S.~K.}\ \bibnamefont {Banerjee}},\ }\href
  {\doibase 10.1103/PhysRevB.82.115426} {\bibfield  {journal} {\bibinfo
  {journal} {Physical Review B}\ }\textbf {\bibinfo {volume} {82}},\ \bibinfo
  {pages} {115426} (\bibinfo {year} {2010})}\BibitemShut {NoStop}%
\bibitem [{\citenamefont {Sahu}\ \emph {et~al.}(2008)\citenamefont {Sahu},
  \citenamefont {Min}, \citenamefont {MacDonald},\ and\ \citenamefont
  {Banerjee}}]{Sahu2008}%
  \BibitemOpen
  \bibfield  {author} {\bibinfo {author} {\bibfnamefont {B.}~\bibnamefont
  {Sahu}}, \bibinfo {author} {\bibfnamefont {H.}~\bibnamefont {Min}}, \bibinfo
  {author} {\bibfnamefont {A.}~\bibnamefont {MacDonald}}, \ and\ \bibinfo
  {author} {\bibfnamefont {S.}~\bibnamefont {Banerjee}},\ }\href {\doibase
  10.1103/PhysRevB.78.045404} {\bibfield  {journal} {\bibinfo  {journal}
  {Physical Review B}\ }\textbf {\bibinfo {volume} {78}},\ \bibinfo {pages}
  {045404} (\bibinfo {year} {2008})}\BibitemShut {NoStop}%
\bibitem [{\citenamefont {Yu}\ and\ \citenamefont {Duan}(2013)}]{Yu2013}%
  \BibitemOpen
  \bibfield  {author} {\bibinfo {author} {\bibfnamefont {W.~J.}\ \bibnamefont
  {Yu}}\ and\ \bibinfo {author} {\bibfnamefont {X.}~\bibnamefont {Duan}},\
  }\href {\doibase 10.1038/srep01248} {\bibfield  {journal} {\bibinfo
  {journal} {Scientific reports}\ }\textbf {\bibinfo {volume} {3}},\ \bibinfo
  {pages} {1248} (\bibinfo {year} {2013})}\BibitemShut {NoStop}%
\bibitem [{\citenamefont {Chang}\ \emph {et~al.}(2006)\citenamefont {Chang},
  \citenamefont {Huang}, \citenamefont {Lu}, \citenamefont {Ho}, \citenamefont
  {Li},\ and\ \citenamefont {Lin}}]{Chang2006}%
  \BibitemOpen
  \bibfield  {author} {\bibinfo {author} {\bibfnamefont {C.}~\bibnamefont
  {Chang}}, \bibinfo {author} {\bibfnamefont {Y.}~\bibnamefont {Huang}},
  \bibinfo {author} {\bibfnamefont {C.}~\bibnamefont {Lu}}, \bibinfo {author}
  {\bibfnamefont {J.}~\bibnamefont {Ho}}, \bibinfo {author} {\bibfnamefont
  {T.}~\bibnamefont {Li}}, \ and\ \bibinfo {author} {\bibfnamefont
  {M.}~\bibnamefont {Lin}},\ }\href {\doibase 10.1016/j.carbon.2005.08.009}
  {\bibfield  {journal} {\bibinfo  {journal} {Carbon}\ }\textbf {\bibinfo
  {volume} {44}},\ \bibinfo {pages} {508} (\bibinfo {year} {2006})}\BibitemShut
  {NoStop}%
\bibitem [{\citenamefont {Son}\ \emph {et~al.}(2006{\natexlab{a}})\citenamefont
  {Son}, \citenamefont {Cohen},\ and\ \citenamefont {Louie}}]{Son2006}%
  \BibitemOpen
  \bibfield  {author} {\bibinfo {author} {\bibfnamefont {Y.-W.}\ \bibnamefont
  {Son}}, \bibinfo {author} {\bibfnamefont {M.~L.}\ \bibnamefont {Cohen}}, \
  and\ \bibinfo {author} {\bibfnamefont {S.~G.}\ \bibnamefont {Louie}},\ }\href
  {\doibase 10.1038/nature05180} {\bibfield  {journal} {\bibinfo  {journal}
  {Nature}\ }\textbf {\bibinfo {volume} {444}},\ \bibinfo {pages} {347}
  (\bibinfo {year} {2006}{\natexlab{a}})}\BibitemShut {NoStop}%
\bibitem [{\citenamefont {Huang}\ \emph {et~al.}(2008)\citenamefont {Huang},
  \citenamefont {Chang},\ and\ \citenamefont {Lin}}]{Huang2008}%
  \BibitemOpen
  \bibfield  {author} {\bibinfo {author} {\bibfnamefont {Y.~C.}\ \bibnamefont
  {Huang}}, \bibinfo {author} {\bibfnamefont {C.~P.}\ \bibnamefont {Chang}}, \
  and\ \bibinfo {author} {\bibfnamefont {M.~F.}\ \bibnamefont {Lin}},\ }\href
  {\doibase 10.1063/1.3028271} {\bibfield  {journal} {\bibinfo  {journal}
  {Journal of Applied Physics}\ }\textbf {\bibinfo {volume} {104}},\ \bibinfo
  {pages} {103714} (\bibinfo {year} {2008})}\BibitemShut {NoStop}%
\bibitem [{\citenamefont {Cai}\ \emph {et~al.}(2010)\citenamefont {Cai},
  \citenamefont {Ruffieux}, \citenamefont {Jaafar}, \citenamefont {Bieri},
  \citenamefont {Braun}, \citenamefont {Blankenburg}, \citenamefont {Muoth},
  \citenamefont {Seitsonen}, \citenamefont {Saleh}, \citenamefont {Feng},
  \citenamefont {M\"{u}llen},\ and\ \citenamefont {Fasel}}]{Cai2010}%
  \BibitemOpen
  \bibfield  {author} {\bibinfo {author} {\bibfnamefont {J.}~\bibnamefont
  {Cai}}, \bibinfo {author} {\bibfnamefont {P.}~\bibnamefont {Ruffieux}},
  \bibinfo {author} {\bibfnamefont {R.}~\bibnamefont {Jaafar}}, \bibinfo
  {author} {\bibfnamefont {M.}~\bibnamefont {Bieri}}, \bibinfo {author}
  {\bibfnamefont {T.}~\bibnamefont {Braun}}, \bibinfo {author} {\bibfnamefont
  {S.}~\bibnamefont {Blankenburg}}, \bibinfo {author} {\bibfnamefont
  {M.}~\bibnamefont {Muoth}}, \bibinfo {author} {\bibfnamefont {A.~P.}\
  \bibnamefont {Seitsonen}}, \bibinfo {author} {\bibfnamefont {M.}~\bibnamefont
  {Saleh}}, \bibinfo {author} {\bibfnamefont {X.}~\bibnamefont {Feng}},
  \bibinfo {author} {\bibfnamefont {K.}~\bibnamefont {M\"{u}llen}}, \ and\
  \bibinfo {author} {\bibfnamefont {R.}~\bibnamefont {Fasel}},\ }\href
  {\doibase 10.1038/nature09211} {\bibfield  {journal} {\bibinfo  {journal}
  {Nature}\ }\textbf {\bibinfo {volume} {466}},\ \bibinfo {pages} {470}
  (\bibinfo {year} {2010})}\BibitemShut {NoStop}%
\bibitem [{\citenamefont {Vo}\ \emph {et~al.}(2014)\citenamefont {Vo},
  \citenamefont {Shekhirev}, \citenamefont {Kunkel}, \citenamefont {Morton},
  \citenamefont {Berglund}, \citenamefont {Kong}, \citenamefont {Wilson},
  \citenamefont {Dowben}, \citenamefont {Enders},\ and\ \citenamefont
  {Sinitskii}}]{Vo2014}%
  \BibitemOpen
  \bibfield  {author} {\bibinfo {author} {\bibfnamefont {T.~H.}\ \bibnamefont
  {Vo}}, \bibinfo {author} {\bibfnamefont {M.}~\bibnamefont {Shekhirev}},
  \bibinfo {author} {\bibfnamefont {D.~A.}\ \bibnamefont {Kunkel}}, \bibinfo
  {author} {\bibfnamefont {M.~D.}\ \bibnamefont {Morton}}, \bibinfo {author}
  {\bibfnamefont {E.}~\bibnamefont {Berglund}}, \bibinfo {author}
  {\bibfnamefont {L.}~\bibnamefont {Kong}}, \bibinfo {author} {\bibfnamefont
  {P.~M.}\ \bibnamefont {Wilson}}, \bibinfo {author} {\bibfnamefont {P.~A.}\
  \bibnamefont {Dowben}}, \bibinfo {author} {\bibfnamefont {A.}~\bibnamefont
  {Enders}}, \ and\ \bibinfo {author} {\bibfnamefont {A.}~\bibnamefont
  {Sinitskii}},\ }\href {\doibase 10.1038/ncomms4189} {\bibfield  {journal}
  {\bibinfo  {journal} {Nature communications}\ }\textbf {\bibinfo {volume}
  {5}},\ \bibinfo {pages} {3189} (\bibinfo {year} {2014})}\BibitemShut
  {NoStop}%
\bibitem [{\citenamefont {Narita}\ \emph {et~al.}(2014)\citenamefont {Narita},
  \citenamefont {Feng}, \citenamefont {Hernandez}, \citenamefont {Jensen},
  \citenamefont {Bonn}, \citenamefont {Yang}, \citenamefont {Verzhbitskiy},
  \citenamefont {Casiraghi}, \citenamefont {Hansen}, \citenamefont {Koch},
  \citenamefont {Fytas}, \citenamefont {Ivasenko}, \citenamefont {Li},
  \citenamefont {Mali}, \citenamefont {Balandina}, \citenamefont {Mahesh},
  \citenamefont {{De Feyter}},\ and\ \citenamefont {M\"{u}llen}}]{Narita2014}%
  \BibitemOpen
  \bibfield  {author} {\bibinfo {author} {\bibfnamefont {A.}~\bibnamefont
  {Narita}}, \bibinfo {author} {\bibfnamefont {X.}~\bibnamefont {Feng}},
  \bibinfo {author} {\bibfnamefont {Y.}~\bibnamefont {Hernandez}}, \bibinfo
  {author} {\bibfnamefont {S.~r.~A.}\ \bibnamefont {Jensen}}, \bibinfo {author}
  {\bibfnamefont {M.}~\bibnamefont {Bonn}}, \bibinfo {author} {\bibfnamefont
  {H.}~\bibnamefont {Yang}}, \bibinfo {author} {\bibfnamefont {I.~A.}\
  \bibnamefont {Verzhbitskiy}}, \bibinfo {author} {\bibfnamefont
  {C.}~\bibnamefont {Casiraghi}}, \bibinfo {author} {\bibfnamefont {M.~R.}\
  \bibnamefont {Hansen}}, \bibinfo {author} {\bibfnamefont {A.~H.~R.}\
  \bibnamefont {Koch}}, \bibinfo {author} {\bibfnamefont {G.}~\bibnamefont
  {Fytas}}, \bibinfo {author} {\bibfnamefont {O.}~\bibnamefont {Ivasenko}},
  \bibinfo {author} {\bibfnamefont {B.}~\bibnamefont {Li}}, \bibinfo {author}
  {\bibfnamefont {K.~S.}\ \bibnamefont {Mali}}, \bibinfo {author}
  {\bibfnamefont {T.}~\bibnamefont {Balandina}}, \bibinfo {author}
  {\bibfnamefont {S.}~\bibnamefont {Mahesh}}, \bibinfo {author} {\bibfnamefont
  {S.}~\bibnamefont {{De Feyter}}}, \ and\ \bibinfo {author} {\bibfnamefont
  {K.}~\bibnamefont {M\"{u}llen}},\ }\href {\doibase 10.1038/nchem.1819}
  {\bibfield  {journal} {\bibinfo  {journal} {Nature chemistry}\ }\textbf
  {\bibinfo {volume} {6}},\ \bibinfo {pages} {126} (\bibinfo {year}
  {2014})}\BibitemShut {NoStop}%
\bibitem [{\citenamefont {Cai}\ \emph {et~al.}(2014)\citenamefont {Cai},
  \citenamefont {Pignedoli}, \citenamefont {Talirz}, \citenamefont {Ruffieux},
  \citenamefont {S\"{o}de}, \citenamefont {Liang}, \citenamefont {Meunier},
  \citenamefont {Berger}, \citenamefont {Li}, \citenamefont {Feng},
  \citenamefont {M\"{u}llen},\ and\ \citenamefont {Fasel}}]{Cai2014}%
  \BibitemOpen
  \bibfield  {author} {\bibinfo {author} {\bibfnamefont {J.}~\bibnamefont
  {Cai}}, \bibinfo {author} {\bibfnamefont {C.~A.}\ \bibnamefont {Pignedoli}},
  \bibinfo {author} {\bibfnamefont {L.}~\bibnamefont {Talirz}}, \bibinfo
  {author} {\bibfnamefont {P.}~\bibnamefont {Ruffieux}}, \bibinfo {author}
  {\bibfnamefont {H.}~\bibnamefont {S\"{o}de}}, \bibinfo {author}
  {\bibfnamefont {L.}~\bibnamefont {Liang}}, \bibinfo {author} {\bibfnamefont
  {V.}~\bibnamefont {Meunier}}, \bibinfo {author} {\bibfnamefont
  {R.}~\bibnamefont {Berger}}, \bibinfo {author} {\bibfnamefont
  {R.}~\bibnamefont {Li}}, \bibinfo {author} {\bibfnamefont {X.}~\bibnamefont
  {Feng}}, \bibinfo {author} {\bibfnamefont {K.}~\bibnamefont {M\"{u}llen}}, \
  and\ \bibinfo {author} {\bibfnamefont {R.}~\bibnamefont {Fasel}},\ }\href
  {\doibase 10.1038/nnano.2014.184} {\bibfield  {journal} {\bibinfo  {journal}
  {Nature Nanotechnology}\ }\textbf {\bibinfo {volume} {9}},\ \bibinfo {pages}
  {896} (\bibinfo {year} {2014})}\BibitemShut {NoStop}%
\bibitem [{\citenamefont {{Costa Gir\~{a}o}}\ \emph {et~al.}(2011)\citenamefont
  {{Costa Gir\~{a}o}}, \citenamefont {Liang}, \citenamefont {Cruz-Silva},
  \citenamefont {Filho},\ and\ \citenamefont {Meunier}}]{CostaGirao2011}%
  \BibitemOpen
  \bibfield  {author} {\bibinfo {author} {\bibfnamefont {E.}~\bibnamefont
  {{Costa Gir\~{a}o}}}, \bibinfo {author} {\bibfnamefont {L.}~\bibnamefont
  {Liang}}, \bibinfo {author} {\bibfnamefont {E.}~\bibnamefont {Cruz-Silva}},
  \bibinfo {author} {\bibfnamefont {A.~G.~S.}\ \bibnamefont {Filho}}, \ and\
  \bibinfo {author} {\bibfnamefont {V.}~\bibnamefont {Meunier}},\ }\href
  {\doibase 10.1103/PhysRevLett.107.135501} {\bibfield  {journal} {\bibinfo
  {journal} {Physical Review Letters}\ }\textbf {\bibinfo {volume} {107}},\
  \bibinfo {pages} {135501} (\bibinfo {year} {2011})}\BibitemShut {NoStop}%
\bibitem [{\citenamefont {Gir\~{a}o}\ \emph {et~al.}(2012)\citenamefont
  {Gir\~{a}o}, \citenamefont {Cruz-Silva},\ and\ \citenamefont
  {Meunier}}]{Girao2012}%
  \BibitemOpen
  \bibfield  {author} {\bibinfo {author} {\bibfnamefont {E.~C.}\ \bibnamefont
  {Gir\~{a}o}}, \bibinfo {author} {\bibfnamefont {E.}~\bibnamefont
  {Cruz-Silva}}, \ and\ \bibinfo {author} {\bibfnamefont {V.}~\bibnamefont
  {Meunier}},\ }\href {\doibase 10.1021/nn302259f} {\bibfield  {journal}
  {\bibinfo  {journal} {ACS nano}\ }\textbf {\bibinfo {volume} {6}},\ \bibinfo
  {pages} {6483} (\bibinfo {year} {2012})}\BibitemShut {NoStop}%
\bibitem [{\citenamefont {Huang}\ \emph {et~al.}(2011)\citenamefont {Huang},
  \citenamefont {Wang},\ and\ \citenamefont {Liang}}]{Huang2011}%
  \BibitemOpen
  \bibfield  {author} {\bibinfo {author} {\bibfnamefont {W.}~\bibnamefont
  {Huang}}, \bibinfo {author} {\bibfnamefont {J.-S.}\ \bibnamefont {Wang}}, \
  and\ \bibinfo {author} {\bibfnamefont {G.}~\bibnamefont {Liang}},\ }\href
  {\doibase 10.1103/PhysRevB.84.045410} {\bibfield  {journal} {\bibinfo
  {journal} {Physical Review B}\ }\textbf {\bibinfo {volume} {84}},\ \bibinfo
  {pages} {045410} (\bibinfo {year} {2011})}\BibitemShut {NoStop}%
\bibitem [{\citenamefont {Liang}\ \emph {et~al.}(2013)\citenamefont {Liang},
  \citenamefont {Gir\~{a}o},\ and\ \citenamefont {Meunier}}]{Liang2013a}%
  \BibitemOpen
  \bibfield  {author} {\bibinfo {author} {\bibfnamefont {L.}~\bibnamefont
  {Liang}}, \bibinfo {author} {\bibfnamefont {E.~C.}\ \bibnamefont
  {Gir\~{a}o}}, \ and\ \bibinfo {author} {\bibfnamefont {V.}~\bibnamefont
  {Meunier}},\ }\href {\doibase 10.1103/PhysRevB.88.035420} {\bibfield
  {journal} {\bibinfo  {journal} {Physical Review B}\ }\textbf {\bibinfo
  {volume} {88}},\ \bibinfo {pages} {035420} (\bibinfo {year}
  {2013})}\BibitemShut {NoStop}%
\bibitem [{\citenamefont {Liang}\ and\ \citenamefont
  {Meunier}(2012)}]{Liang2012}%
  \BibitemOpen
  \bibfield  {author} {\bibinfo {author} {\bibfnamefont {L.}~\bibnamefont
  {Liang}}\ and\ \bibinfo {author} {\bibfnamefont {V.}~\bibnamefont
  {Meunier}},\ }\href {\doibase 10.1103/PhysRevB.86.195404} {\bibfield
  {journal} {\bibinfo  {journal} {Physical Review B}\ }\textbf {\bibinfo
  {volume} {86}},\ \bibinfo {pages} {195404} (\bibinfo {year}
  {2012})}\BibitemShut {NoStop}%
\bibitem [{\citenamefont {Liang}\ and\ \citenamefont
  {Meunier}(2013)}]{Liang2013}%
  \BibitemOpen
  \bibfield  {author} {\bibinfo {author} {\bibfnamefont {L.}~\bibnamefont
  {Liang}}\ and\ \bibinfo {author} {\bibfnamefont {V.}~\bibnamefont
  {Meunier}},\ }\href {\doibase 10.1063/1.4800777} {\bibfield  {journal}
  {\bibinfo  {journal} {Applied Physics Letters}\ }\textbf {\bibinfo {volume}
  {102}},\ \bibinfo {pages} {143101} (\bibinfo {year} {2013})}\BibitemShut
  {NoStop}%
\bibitem [{\citenamefont {Wu}\ and\ \citenamefont {Zeng}(2008)}]{Wu2008}%
  \BibitemOpen
  \bibfield  {author} {\bibinfo {author} {\bibfnamefont {X.}~\bibnamefont
  {Wu}}\ and\ \bibinfo {author} {\bibfnamefont {X.~C.}\ \bibnamefont {Zeng}},\
  }\href {\doibase 10.1007/s12274-008-8001-z} {\bibfield  {journal} {\bibinfo
  {journal} {Nano Research}\ }\textbf {\bibinfo {volume} {1}},\ \bibinfo
  {pages} {40} (\bibinfo {year} {2008})}\BibitemShut {NoStop}%
\bibitem [{\citenamefont {Sevin\c{c}li}\ \emph {et~al.}(2008)\citenamefont
  {Sevin\c{c}li}, \citenamefont {Topsakal},\ and\ \citenamefont
  {Ciraci}}]{Sevincli2008b}%
  \BibitemOpen
  \bibfield  {author} {\bibinfo {author} {\bibfnamefont {H.}~\bibnamefont
  {Sevin\c{c}li}}, \bibinfo {author} {\bibfnamefont {M.}~\bibnamefont
  {Topsakal}}, \ and\ \bibinfo {author} {\bibfnamefont {S.}~\bibnamefont
  {Ciraci}},\ }\href {\doibase 10.1103/PhysRevB.78.245402} {\bibfield
  {journal} {\bibinfo  {journal} {Physical Review B}\ }\textbf {\bibinfo
  {volume} {78}},\ \bibinfo {pages} {245402} (\bibinfo {year} {2008})},\
  \Eprint {http://arxiv.org/abs/arXiv:0711.2414v3} {arXiv:arXiv:0711.2414v3}
  \BibitemShut {NoStop}%
\bibitem [{\citenamefont {Ihnatsenka}\ \emph {et~al.}(2009)\citenamefont
  {Ihnatsenka}, \citenamefont {Zozoulenko},\ and\ \citenamefont
  {Kirczenow}}]{Ihnatsenka2009}%
  \BibitemOpen
  \bibfield  {author} {\bibinfo {author} {\bibfnamefont {S.}~\bibnamefont
  {Ihnatsenka}}, \bibinfo {author} {\bibfnamefont {I.}~\bibnamefont
  {Zozoulenko}}, \ and\ \bibinfo {author} {\bibfnamefont {G.}~\bibnamefont
  {Kirczenow}},\ }\href {\doibase 10.1103/PhysRevB.80.155415} {\bibfield
  {journal} {\bibinfo  {journal} {Physical Review B}\ }\textbf {\bibinfo
  {volume} {80}},\ \bibinfo {pages} {155415} (\bibinfo {year}
  {2009})}\BibitemShut {NoStop}%
\bibitem [{\citenamefont {Son}\ \emph {et~al.}(2006{\natexlab{b}})\citenamefont
  {Son}, \citenamefont {Cohen},\ and\ \citenamefont
  {Louie}}]{PhysRevLett.97.216803}%
  \BibitemOpen
  \bibfield  {author} {\bibinfo {author} {\bibfnamefont {Y.-W.}\ \bibnamefont
  {Son}}, \bibinfo {author} {\bibfnamefont {M.~L.}\ \bibnamefont {Cohen}}, \
  and\ \bibinfo {author} {\bibfnamefont {S.~G.}\ \bibnamefont {Louie}},\ }\href
  {\doibase 10.1103/PhysRevLett.97.216803} {\bibfield  {journal} {\bibinfo
  {journal} {Physical Review Letters}\ }\textbf {\bibinfo {volume} {97}},\
  \bibinfo {pages} {216803} (\bibinfo {year} {2006}{\natexlab{b}})}\BibitemShut
  {NoStop}%
\bibitem [{\citenamefont {White}\ \emph {et~al.}(2007)\citenamefont {White},
  \citenamefont {Li}, \citenamefont {Gunlycke},\ and\ \citenamefont
  {Mintmire}}]{White2007}%
  \BibitemOpen
  \bibfield  {author} {\bibinfo {author} {\bibfnamefont {C.~T.}\ \bibnamefont
  {White}}, \bibinfo {author} {\bibfnamefont {J.}~\bibnamefont {Li}}, \bibinfo
  {author} {\bibfnamefont {D.}~\bibnamefont {Gunlycke}}, \ and\ \bibinfo
  {author} {\bibfnamefont {J.~W.}\ \bibnamefont {Mintmire}},\ }\href {\doibase
  10.1021/nl0627745} {\bibfield  {journal} {\bibinfo  {journal} {Nano letters}\
  }\textbf {\bibinfo {volume} {7}},\ \bibinfo {pages} {825} (\bibinfo {year}
  {2007})}\BibitemShut {NoStop}%
\bibitem [{\citenamefont {Gunlycke}\ and\ \citenamefont
  {White}(2008)}]{Gunlycke2008}%
  \BibitemOpen
  \bibfield  {author} {\bibinfo {author} {\bibfnamefont {D.}~\bibnamefont
  {Gunlycke}}\ and\ \bibinfo {author} {\bibfnamefont {C.}~\bibnamefont
  {White}},\ }\href {\doibase 10.1103/PhysRevB.77.115116} {\bibfield  {journal}
  {\bibinfo  {journal} {Physical Review B}\ }\textbf {\bibinfo {volume} {77}},\
  \bibinfo {pages} {115116} (\bibinfo {year} {2008})}\BibitemShut {NoStop}%
\bibitem [{\citenamefont {Partoens}\ and\ \citenamefont
  {Peeters}(2006)}]{Partoens2006}%
  \BibitemOpen
  \bibfield  {author} {\bibinfo {author} {\bibfnamefont {B.}~\bibnamefont
  {Partoens}}\ and\ \bibinfo {author} {\bibfnamefont {F.}~\bibnamefont
  {Peeters}},\ }\href {\doibase 10.1103/PhysRevB.74.075404} {\bibfield
  {journal} {\bibinfo  {journal} {Physical Review B}\ }\textbf {\bibinfo
  {volume} {74}},\ \bibinfo {pages} {075404} (\bibinfo {year}
  {2006})}\BibitemShut {NoStop}%
\bibitem [{\citenamefont {Tada}\ and\ \citenamefont
  {Watanabe}(2002)}]{Tada2002}%
  \BibitemOpen
  \bibfield  {author} {\bibinfo {author} {\bibfnamefont {K.}~\bibnamefont
  {Tada}}\ and\ \bibinfo {author} {\bibfnamefont {K.}~\bibnamefont
  {Watanabe}},\ }\href {\doibase 10.1103/PhysRevLett.88.127601} {\bibfield
  {journal} {\bibinfo  {journal} {Physical Review Letters}\ }\textbf {\bibinfo
  {volume} {88}},\ \bibinfo {pages} {127601} (\bibinfo {year}
  {2002})}\BibitemShut {NoStop}%
\bibitem [{\citenamefont {Araidai}\ \emph {et~al.}(2004)\citenamefont
  {Araidai}, \citenamefont {Nakamura},\ and\ \citenamefont
  {Watanabe}}]{Araidai2004}%
  \BibitemOpen
  \bibfield  {author} {\bibinfo {author} {\bibfnamefont {M.}~\bibnamefont
  {Araidai}}, \bibinfo {author} {\bibfnamefont {Y.}~\bibnamefont {Nakamura}}, \
  and\ \bibinfo {author} {\bibfnamefont {K.}~\bibnamefont {Watanabe}},\ }\href
  {\doibase 10.1103/PhysRevB.70.245410} {\bibfield  {journal} {\bibinfo
  {journal} {Physical Review B}\ }\textbf {\bibinfo {volume} {70}},\ \bibinfo
  {pages} {245410} (\bibinfo {year} {2004})}\BibitemShut {NoStop}%
\bibitem [{\citenamefont {DiMaria}\ \emph {et~al.}(1993)\citenamefont
  {DiMaria}, \citenamefont {Cartier},\ and\ \citenamefont
  {Arnold}}]{DiMaria1993}%
  \BibitemOpen
  \bibfield  {author} {\bibinfo {author} {\bibfnamefont {D.~J.}\ \bibnamefont
  {DiMaria}}, \bibinfo {author} {\bibfnamefont {E.}~\bibnamefont {Cartier}}, \
  and\ \bibinfo {author} {\bibfnamefont {D.}~\bibnamefont {Arnold}},\ }\href
  {\doibase 10.1063/1.352936} {\bibfield  {journal} {\bibinfo  {journal}
  {Journal of Applied Physics}\ }\textbf {\bibinfo {volume} {73}},\ \bibinfo
  {pages} {3367} (\bibinfo {year} {1993})}\BibitemShut {NoStop}%
\bibitem [{\citenamefont {Kibis}\ \emph
  {et~al.}(2005{\natexlab{a}})\citenamefont {Kibis}, \citenamefont
  {Malevannyy}, \citenamefont {Huggett}, \citenamefont {Parfitt},\ and\
  \citenamefont {Portnoi}}]{Kibis2005a}%
  \BibitemOpen
  \bibfield  {author} {\bibinfo {author} {\bibfnamefont {O.~V.}\ \bibnamefont
  {Kibis}}, \bibinfo {author} {\bibfnamefont {S.~V.}\ \bibnamefont
  {Malevannyy}}, \bibinfo {author} {\bibfnamefont {L.}~\bibnamefont {Huggett}},
  \bibinfo {author} {\bibfnamefont {D.~G.~W.}\ \bibnamefont {Parfitt}}, \ and\
  \bibinfo {author} {\bibfnamefont {M.~E.}\ \bibnamefont {Portnoi}},\ }\href
  {\doibase 10.1080/02726340590957416} {\bibfield  {journal} {\bibinfo
  {journal} {Electromagnetics}\ }\textbf {\bibinfo {volume} {25}},\ \bibinfo
  {pages} {425} (\bibinfo {year} {2005}{\natexlab{a}})}\BibitemShut {NoStop}%
\bibitem [{\citenamefont {Kibis}\ \emph
  {et~al.}(2005{\natexlab{b}})\citenamefont {Kibis}, \citenamefont {Parfitt},\
  and\ \citenamefont {Portnoi}}]{Kibis2005}%
  \BibitemOpen
  \bibfield  {author} {\bibinfo {author} {\bibfnamefont {O.}~\bibnamefont
  {Kibis}}, \bibinfo {author} {\bibfnamefont {D.}~\bibnamefont {Parfitt}}, \
  and\ \bibinfo {author} {\bibfnamefont {M.}~\bibnamefont {Portnoi}},\ }\href
  {\doibase 10.1103/PhysRevB.71.035411} {\bibfield  {journal} {\bibinfo
  {journal} {Physical Review B}\ }\textbf {\bibinfo {volume} {71}},\ \bibinfo
  {pages} {035411} (\bibinfo {year} {2005}{\natexlab{b}})}\BibitemShut
  {NoStop}%
\bibitem [{\citenamefont {Batrakov}\ \emph {et~al.}(2010)\citenamefont
  {Batrakov}, \citenamefont {Kibis}, \citenamefont {Kuzhir}, \citenamefont
  {{Rosenau da Costa}},\ and\ \citenamefont {Portnoi}}]{Batrakov2010}%
  \BibitemOpen
  \bibfield  {author} {\bibinfo {author} {\bibfnamefont {K.~G.}\ \bibnamefont
  {Batrakov}}, \bibinfo {author} {\bibfnamefont {O.~V.}\ \bibnamefont {Kibis}},
  \bibinfo {author} {\bibfnamefont {P.~P.}\ \bibnamefont {Kuzhir}}, \bibinfo
  {author} {\bibfnamefont {M.}~\bibnamefont {{Rosenau da Costa}}}, \ and\
  \bibinfo {author} {\bibfnamefont {M.~E.}\ \bibnamefont {Portnoi}},\ }\href
  {\doibase 10.1117/1.3436585} {\bibfield  {journal} {\bibinfo  {journal}
  {Journal of Nanophotonics}\ }\textbf {\bibinfo {volume} {4}},\ \bibinfo
  {pages} {041665} (\bibinfo {year} {2010})}\BibitemShut {NoStop}%
\bibitem [{\citenamefont {Fujita}\ \emph {et~al.}(1996)\citenamefont {Fujita},
  \citenamefont {Wakabayashi}, \citenamefont {Nakada},\ and\ \citenamefont
  {Kusakabe}}]{Fujita1996}%
  \BibitemOpen
  \bibfield  {author} {\bibinfo {author} {\bibfnamefont {M.}~\bibnamefont
  {Fujita}}, \bibinfo {author} {\bibfnamefont {K.}~\bibnamefont {Wakabayashi}},
  \bibinfo {author} {\bibfnamefont {K.}~\bibnamefont {Nakada}}, \ and\ \bibinfo
  {author} {\bibfnamefont {K.}~\bibnamefont {Kusakabe}},\ }\href {\doibase
  10.1143/JPSJ.65.1920} {\bibfield  {journal} {\bibinfo  {journal} {Journal of
  the Physics Society Japan}\ }\textbf {\bibinfo {volume} {65}},\ \bibinfo
  {pages} {1920} (\bibinfo {year} {1996})}\BibitemShut {NoStop}%
\bibitem [{\citenamefont {Rotkin}\ and\ \citenamefont
  {Hess}(2004)}]{Rotkin2004}%
  \BibitemOpen
  \bibfield  {author} {\bibinfo {author} {\bibfnamefont {S.~V.}\ \bibnamefont
  {Rotkin}}\ and\ \bibinfo {author} {\bibfnamefont {K.}~\bibnamefont {Hess}},\
  }\href {\doibase 10.1063/1.1710717} {\bibfield  {journal} {\bibinfo
  {journal} {Applied Physics Letters}\ }\textbf {\bibinfo {volume} {84}},\
  \bibinfo {pages} {3139} (\bibinfo {year} {2004})}\BibitemShut {NoStop}%
\bibitem [{\citenamefont {Wei}\ \emph {et~al.}(2010)\citenamefont {Wei},
  \citenamefont {Wang}, \citenamefont {Kim}, \citenamefont {Kim}, \citenamefont
  {Hu}, \citenamefont {Yakes}, \citenamefont {Laracuente}, \citenamefont {Dai},
  \citenamefont {Marder}, \citenamefont {Berger}, \citenamefont {King},
  \citenamefont {de~Heer}, \citenamefont {Sheehan},\ and\ \citenamefont
  {Riedo}}]{Wei2010}%
  \BibitemOpen
  \bibfield  {author} {\bibinfo {author} {\bibfnamefont {Z.}~\bibnamefont
  {Wei}}, \bibinfo {author} {\bibfnamefont {D.}~\bibnamefont {Wang}}, \bibinfo
  {author} {\bibfnamefont {S.}~\bibnamefont {Kim}}, \bibinfo {author}
  {\bibfnamefont {S.-Y.}\ \bibnamefont {Kim}}, \bibinfo {author} {\bibfnamefont
  {Y.}~\bibnamefont {Hu}}, \bibinfo {author} {\bibfnamefont {M.~K.}\
  \bibnamefont {Yakes}}, \bibinfo {author} {\bibfnamefont {A.~R.}\ \bibnamefont
  {Laracuente}}, \bibinfo {author} {\bibfnamefont {Z.}~\bibnamefont {Dai}},
  \bibinfo {author} {\bibfnamefont {S.~R.}\ \bibnamefont {Marder}}, \bibinfo
  {author} {\bibfnamefont {C.}~\bibnamefont {Berger}}, \bibinfo {author}
  {\bibfnamefont {W.~P.}\ \bibnamefont {King}}, \bibinfo {author}
  {\bibfnamefont {W.~a.}\ \bibnamefont {de~Heer}}, \bibinfo {author}
  {\bibfnamefont {P.~E.}\ \bibnamefont {Sheehan}}, \ and\ \bibinfo {author}
  {\bibfnamefont {E.}~\bibnamefont {Riedo}},\ }\href {\doibase
  10.1126/science.1188119} {\bibfield  {journal} {\bibinfo  {journal} {Science
  (New York, N.Y.)}\ }\textbf {\bibinfo {volume} {328}},\ \bibinfo {pages}
  {1373} (\bibinfo {year} {2010})}\BibitemShut {NoStop}%
\bibitem [{\citenamefont {Liu}\ \emph {et~al.}(2013)\citenamefont {Liu},
  \citenamefont {Ma}, \citenamefont {Shi}, \citenamefont {Zhou}, \citenamefont
  {Gong}, \citenamefont {Lei}, \citenamefont {Yang}, \citenamefont {Zhang},
  \citenamefont {Yu}, \citenamefont {Hackenberg}, \citenamefont {Babakhani},
  \citenamefont {Idrobo}, \citenamefont {Vajtai}, \citenamefont {Lou},\ and\
  \citenamefont {Ajayan}}]{Liu2013a}%
  \BibitemOpen
  \bibfield  {author} {\bibinfo {author} {\bibfnamefont {Z.}~\bibnamefont
  {Liu}}, \bibinfo {author} {\bibfnamefont {L.}~\bibnamefont {Ma}}, \bibinfo
  {author} {\bibfnamefont {G.}~\bibnamefont {Shi}}, \bibinfo {author}
  {\bibfnamefont {W.}~\bibnamefont {Zhou}}, \bibinfo {author} {\bibfnamefont
  {Y.}~\bibnamefont {Gong}}, \bibinfo {author} {\bibfnamefont {S.}~\bibnamefont
  {Lei}}, \bibinfo {author} {\bibfnamefont {X.}~\bibnamefont {Yang}}, \bibinfo
  {author} {\bibfnamefont {J.}~\bibnamefont {Zhang}}, \bibinfo {author}
  {\bibfnamefont {J.}~\bibnamefont {Yu}}, \bibinfo {author} {\bibfnamefont
  {K.~P.}\ \bibnamefont {Hackenberg}}, \bibinfo {author} {\bibfnamefont
  {A.}~\bibnamefont {Babakhani}}, \bibinfo {author} {\bibfnamefont {J.-C.}\
  \bibnamefont {Idrobo}}, \bibinfo {author} {\bibfnamefont {R.}~\bibnamefont
  {Vajtai}}, \bibinfo {author} {\bibfnamefont {J.}~\bibnamefont {Lou}}, \ and\
  \bibinfo {author} {\bibfnamefont {P.~M.}\ \bibnamefont {Ajayan}},\ }\href
  {\doibase 10.1038/nnano.2012.256} {\bibfield  {journal} {\bibinfo  {journal}
  {Nature nanotechnology}\ }\textbf {\bibinfo {volume} {8}},\ \bibinfo {pages}
  {119} (\bibinfo {year} {2013})}\BibitemShut {NoStop}%
\bibitem [{\citenamefont {Ribas}\ \emph {et~al.}(2011)\citenamefont {Ribas},
  \citenamefont {Singh}, \citenamefont {Sorokin},\ and\ \citenamefont
  {Yakobson}}]{Ribas2010}%
  \BibitemOpen
  \bibfield  {author} {\bibinfo {author} {\bibfnamefont {M.~A.}\ \bibnamefont
  {Ribas}}, \bibinfo {author} {\bibfnamefont {A.~K.}\ \bibnamefont {Singh}},
  \bibinfo {author} {\bibfnamefont {P.~B.}\ \bibnamefont {Sorokin}}, \ and\
  \bibinfo {author} {\bibfnamefont {B.~I.}\ \bibnamefont {Yakobson}},\ }\href
  {\doibase 10.1007/s12274-010-0084-7} {\bibfield  {journal} {\bibinfo
  {journal} {Nano Research}\ }\textbf {\bibinfo {volume} {4}},\ \bibinfo
  {pages} {143} (\bibinfo {year} {2011})}\BibitemShut {NoStop}%
\bibitem [{\citenamefont {Magda}\ \emph {et~al.}(2014)\citenamefont {Magda},
  \citenamefont {Jin}, \citenamefont {Hagym\'{a}si}, \citenamefont
  {Vancs\'{o}}, \citenamefont {Osv\'{a}th}, \citenamefont {Nemes-Incze},
  \citenamefont {Hwang}, \citenamefont {Bir\'{o}},\ and\ \citenamefont
  {Tapaszt\'{o}}}]{Magda2014}%
  \BibitemOpen
  \bibfield  {author} {\bibinfo {author} {\bibfnamefont {G.~Z.}\ \bibnamefont
  {Magda}}, \bibinfo {author} {\bibfnamefont {X.}~\bibnamefont {Jin}}, \bibinfo
  {author} {\bibfnamefont {I.}~\bibnamefont {Hagym\'{a}si}}, \bibinfo {author}
  {\bibfnamefont {P.}~\bibnamefont {Vancs\'{o}}}, \bibinfo {author}
  {\bibfnamefont {Z.}~\bibnamefont {Osv\'{a}th}}, \bibinfo {author}
  {\bibfnamefont {P.}~\bibnamefont {Nemes-Incze}}, \bibinfo {author}
  {\bibfnamefont {C.}~\bibnamefont {Hwang}}, \bibinfo {author} {\bibfnamefont
  {L.~P.}\ \bibnamefont {Bir\'{o}}}, \ and\ \bibinfo {author} {\bibfnamefont
  {L.}~\bibnamefont {Tapaszt\'{o}}},\ }\href {\doibase 10.1038/nature13831}
  {\bibfield  {journal} {\bibinfo  {journal} {Nature}\ }\textbf {\bibinfo
  {volume} {514}},\ \bibinfo {pages} {608} (\bibinfo {year}
  {2014})}\BibitemShut {NoStop}%
\end{thebibliography}%
\end{document}